\documentclass[italian,english,authoryear]{article}
\usepackage{mathpazo}
\usepackage[T1]{fontenc}
\usepackage[utf8]{inputenc}
\usepackage[a4paper]{geometry}
\usepackage{babel}
\usepackage{array}
\usepackage{textcomp}
\usepackage{multirow}
\usepackage{amssymb}
\usepackage{graphicx}
\usepackage[authoryear]{natbib}
\usepackage[unicode=true,pdfusetitle,
 bookmarks=true,bookmarksnumbered=false,bookmarksopen=false,
 breaklinks=false,pdfborder={0 0 0},pdfborderstyle={},backref=false,colorlinks=false]
 {hyperref}

\makeatletter

\newcommand*\LyXHairSpace{\hspace{1pt}}
\newcommand*\LyXThinSpace{\,\hspace{0pt}}
\providecommand{\tabularnewline}{\\}

\newcommand{\lyxaddress}[1]{
	\par {\raggedright #1
	\vspace{1.4em}
	\noindent\par}
}

\@ifundefined{date}{}{\date{}}

\makeatother

\begin{document}
\title{Impacts and risks of ``realistic'' global warming projections for
the 21st century}
\author{Nicola Scafetta}
\maketitle

\lyxaddress{$^{1}$Department of Earth Sciences, Environment and Georesources,
University of Naples Federico II, Complesso Universitario di Monte
S. Angelo, Via Vicinale Cupa Cintia, 21, 80126 Naples (NA), Italy.}
\begin{abstract}
The IPCC AR6 assessment of the impacts and risks associated with projected
climate changes for the 21st century is both alarming and ambiguous.
According to computer projections, global surface may warm from 1.3
to 8.0 °C by 2100, depending on the global climate model (GCM) and
the shared socioeconomic pathway (SSP) scenario used for the simulations.
Actual climate-change hazards are estimated to be high and very high
if the global surface temperature rises, respectively, more than 2.0
°C and 3.0 °C above pre-industrial levels. Recent studies, however,
showed that a substantial number of CMIP6 GCMs run ``\emph{too hot}''
because they appear to be too sensitive to radiative forcing, and
that the high/extreme emission scenarios SSP3-7.0 and SSP5-8.5 are
to be rejected because judged to be \emph{unlikely} and \emph{highly
unlikely}, respectively. Yet, the IPCC AR6 mostly focused on such
alarmistic scenarios for risk assessments. This paper examines the
impacts and risks of ``realistic'' climate change projections for
the 21st century generated by assessing the theoretical models and
integrating them with the existing empirical knowledge on global warming
and the various natural cycles of climate change that have been recorded
by a variety of scientists and historians. This is achieved by combining
the SSP2-4.5 scenario (which is the most likely SSP according to the
current policies reported by the International Energy Agency) and
empirically optimized climate modeling. According to recent research,
the GCM macro-ensemble that best hindcast the global surface warming
observed from 1980--1990 to 2012--2022 should be made up of models
that are characterized by a low equilibrium climate sensitivity (ECS)
($1.5<ECS\lessapprox3.0$ °C), in contrast to the IPCC AR6 \emph{likely}
and \emph{very likely} ECS ranges of 2.5-4.0 °C and 2.0-5.0 °C, respectively.
I show that the low-ECS macro-GCM with the SSP2-4.5 scenario projects
a global surface temperature warming of 1.68-3.09 °C by 2080--2100
instead of1.98-3.82 °C obtained with the GCMs with ECS in the 2.5-4.0
°C range. However, if the global surface temperature records are affected
by significant non-climatic warm biases --- as suggested by satellite-based
lower troposphere temperature records and current studies on urban
heat island effects --- the same climate simulations should be scaled
down by about 30\%, resulting in a warming of about 1.18-2.16 °C by
2080--2100. Furthermore, similar moderate warming estimates (1.15-2.52
°C) are also projected by alternative empirically derived models that
aim to recreate the decadal-to-millennial natural climatic oscillations,
which the GCMs do not reproduce. The proposed methodologies aim to
simulate hypothetical models supposed to optimally hindcast the actual
available data. The obtained climate projections show that the expected
global surface warming for the 21st century will likely be mild, that
is, no more than 2.5-3.0 °C and, on average, likely below the 2.0
°C threshold. This should allow for the mitigation and management
of the most dangerous climate-change related hazards through appropriate
low-cost adaptation policies. In conclusion, enforcing expensive decarbonization
and net-zero emission scenarios, such as SSP1-2.6, is not required
because the Paris Agreement temperature target of keeping global warming
below 2 °C throughout the 21st century should be compatible also with
moderate and pragmatic shared socioeconomic pathways such as the SSP2-4.5.
\end{abstract}
\begin{itemize}
\item \textbf{Keywords:} Climate change; Climate models; Shared socioeconomic
pathways; 21st-century climate projections; Impacts and risks assessment.
\selectlanguage{italian}%
\item \textbf{Cite as:} Scafetta, N.: 2024. Impacts and risks of “realistic”
global warming projections for the 21st century. \emph{Geoscience
Frontiers} 15(2), 101774. (\href{https://doi.org/10.1016/j.gsf.2023.101774}{https://doi.org/10.1016/j.gsf.2023.101774}\foreignlanguage{english}{)}
\end{itemize}
\newpage{}

\textbf{Highlights:}
\begin{itemize}
\item The IPCC AR6 assessment of likely impacts and risks of 21st-century
climate change is highly uncertain.
\item Most climate models, however, run too hot, and the SSP3-7.0 and SSP5-8.5
scenarios are unlikely.
\item New climate change projections for the 21st century were generated
using best-performing climate models, and
\item empirical climate modeling of natural cycles, and calibration on lower
troposphere temperature data.
\item Only the realistic and moderate SSP2-4.5 scenario is considered for
climate-change impact and risk assessments.
\item Net-zero emission policies are not necessary because SSP2-4.5 is sufficient
to limit climate change hazards to manageable levels.
\end{itemize}

\section{Introduction}

The IPCC Sixth Assessment Report (AR6) Working Group II \citep{IPCC_2022a}
assessed that the environmental impacts of climate changes in the
21st century --- such as rising sea levels and temperatures, increased
extreme weather events, droughts and floods, and the spread of wildfires
--- could significantly affect environments, societies and many aspects
of human economy and life. Projected climate changes could also directly
and indirectly impact human health by inducing changes in many environmental
factors associated with infectious and respiratory diseases, changes
in heat and cold-related morbidity and mortality, changes in food
production, changes in sociopolitical tensions and conflicts, and
many others \citep{Rocque_2021}.

In fact, despite CO\textsubscript{2} is, strictly speaking, not a
pollutant, and its atmospheric increase is actually greening the Earth
due to its fertilization effects \citep{Zhu_2016,Piao_2019} and a
warmer climate does also present a number of benefits, today there
is growing concern that, during the 21st century, global warming induced
by anthropogenic greenhouse gas emissions due to the persistent use
of fossil fuels (which mostly emits CO\textsubscript{2}) could exceed
temperature levels that are considered sustainable with the current
adaptive capacity of most communities \citep{IPCC_2018,IPCC_2021}.
It is claimed that dangerous levels of global warming could be exceeded
by as soon as 2050 unless a number of expensive adaptation and mitigation
strategies are implemented such as energy policies attempting to reach
net-zero emissions by about 2050 \citep{European Commission,IPCC_2022b}.

The term \textquotedbl net-zero emissions\textquotedbl{} refers to
achieving a net balance between man-made greenhouse gas (GHG) emissions
produced and GHG removed from the atmosphere. Given the present limited
ability to remove GHGs from the atmosphere, net-zero requires swiftly
limiting the use of fossil fuels - coal, oil, and gas - and transitioning
toward renewable green energy in all sectors of the economy.

Depending on the magnitude and speed of the evolving physical processes,
climate changes may have both beneficial and harmful consequences.
These effects are under the control of complex and still poorly understood
nonlinear dynamical systems. As the temperature continues to warm,
some nations may see immediate advantages, but the long-term consequences
could be harmful, while others may see the opposite. Depending on
elements including geographic location, socioeconomic conditions,
and capacity for adaptation, the distribution of climate-change impacts
and risks also considerably changes from place to place. By aggregating
several indicators, \citet{Tol_2015} estimated that climate change
could globally have a net positive economic welfare impact only if
the global surface warming remains roughly below 2.0 °C above its
1850--1900 pre-industrial levels throughout the 21st century. However,
\citet{Tol_2015} also highlighted the large uncertainty associated
with such claims because, for example, on eleven analyzed economical
risk estimates for a warming of 2.5 °C, it was found that the researchers
disagree on the sign of the net impact: four were positive and seven
negative.

For example, most of the early net gains should occur because increasing
carbon dioxide in the atmosphere reduces water stress in plants, causing
them to grow quicker, and because wealthy countries are mostly concentrated
in temperate climatic zones, which should benefit more from a moderate
warming. However, if the temperature rises too much, the water stress
in the plants could grow. In general, although the situation varies
from place to place, the scientific literature suggests that on a
global scale, even if there may be initial economic net gains from
global warming, such gains may be offset by losses as the environment
continues to warm (particularly above 2.5-3.0 °C) in a too short time
period \citep{IPCC_2022a}. After then, societies will be able to
recover economically only after effectively adapting to the new climatic
conditions.

The Working Group I of the IPCC AR6 \citep{IPCC_2021} stated that
the global surface temperature in the first two decades of the 21st
century (2001--2020) was about 1 {[}0.84 to 1.10{]} °C higher than
the (1850--1900) pre-industrial period. Current computer-based global
climate models (GCMs) suggest that the 1.5 °C warming level could,
on average, be easily exceeded by as early as 2030, and the 2 °C warming
level could be exceeded by about 2050--2060 \citep{IPCC_2018,IPCC_2021}.
Consequently, because the global surface temperature is expected to
rise very fast, climate changes are being considered by the United
Nations a kind of imminent threat, especially for the impoverished
societies that could lack the resources required for quickly cope
and/or adapt to it. Thus, aggressive, and expensive climate policies
aimed at significantly reducing anthropogenic greenhouse gas emissions
from fossil fuels are being advocated for mitigating future climate
changes in annual COPs (Conferences of the Parties) promoted by United
Nations since 1995.

For example, in 2015 at COP-21 in Paris the \citet{UN_Paris_2023}
reached a landmark agreement to combat climate change by promoting
international policies to keep global warming ``\emph{below 2 °C}''
\citep{Gao} while even pursuing ambitious efforts ``\emph{to limit
the temperature increase to 1.5 °C above pre-industrial levels}''
by 2030 by enforcing Net-Zero by 2050 policies. \citet{IPCC_2018}
and \citet{Meinshausen} argued that only by executing the Paris Agreement
pledges of net-zero emissions by 2050 could restrict global surface
warming to less than the ``safe'' threshold of about 2 °C. For similar
reasons, editors of geoscience and health journals have being calling
for urgent collective climate-mitigation actions \citep{McNutt} because
``\emph{our planet is in crisis}'' \citep{Filippelli} and because
the projected global warming for the 21st century would be the ``\emph{greatest
threat to global public health}'' \citep{Atwoli} if not immediately
mitigated as agreed in Paris. However, many severe uncertainties and
concerns on such claims remain.

In fact, fossil fuels continue to supply most of the inexpensive energy
that is today used worldwide, which is allowing developing countries
to quickly raise their living and welfare standards. Indeed, while
wealthy Western Nations seek to phase out coal power plants and, in
general, are attempting to substantially limit the use of fossil fuels
through impractical, expensive and unpopular rapid-decarbonization
policies, the majority of the rest of the world is significantly increasing
their emissions \citep{Crippa}. In particular, China, India, Indonesia,
Japan and Vietnam, among other Asian countries, are constructing about
400 coal-fired power plants and planning to build another 400 units
in the near future \citep{GEM_2023}. Furthermore, the rapid depletion
of critical metals required for a rapid shift from fossil fuel to
renewable green-energies calls the possibility of such a technological
transition into serious doubt \citep{Groves} and, in general, economic
meta-analyses indicate that the costs of implementing the net-zero
policies required to meet the Paris climate targets may outweigh the
benefits even in the worst-case-scenario global warming \citep{Tol2023}.
As a result, there is no political agreement among world nations on
how to manage future climate-change-related hazards.

World nations disagree on the actual threats, and, in fact, they prioritize
different issues, likely because of the scientific uncertainty about
future climate changes, as well as of their potential impacts and
associated risks. The still existing large scientific uncertainty
on this matter may lead to detrimental decisions either if a climatic
alarm is overstated or understated. Policy disparities among countries
would make finding effective solutions to potential climate-change-related
hazards hard, while also risking harmful economic imbalances in the
interactions between and within states. It should be evident to all
that if a few wealthy countries significantly reduce their emissions
while all the others significantly raise them, total emissions may
not fall but, instead, increase and climate change will not be mitigated.
For example, net-zero emission policies are being strongly advocated
in the 27 nations of the European Union that, however, in 2022 contributed
only 6.7\% of the GHG global total emissions \citep{Crippa}; moreover,
for the year 2022 about half of the EU countries -- including Italy
(6.70 t CO\textsubscript{2}eq/cap) and France (6.50 t CO\textsubscript{2}eq/cap)
-- produced GHG per capita emissions less than the world total (6.76
t CO\textsubscript{2}eq/cap).

The potential impacts and risks related to projected climate changes
for the 21st century must be estimated more precisely by constraining
the various existing physical and economic uncertainties. This operation
is necessary to conduct proper cost-benefit analyses of the required
mitigation and/or adaptation policies in the hope that by being less
controversial they could be more likely accepted by all nations. In
fact, either underestimating or exaggerating the alarm of a possible
climate-change crisis could have harmful economic implications that
can negatively impact human economy and life at both global and local
scales.

The main scientific uncertainties surrounding the physics of climate
change are reviewed in Section 2, along with how they could be significantly
constrained by using recent advances in scientific knowledge. Realistic
projections of climate changes for the 21st century and their associated
impacts and risks are evaluated in Section 4. Finally, Section 5 proposes
an alternative interpretation based on empirical climate change modeling
for the 21st century that takes into account the implications of additional
physical uncertainties concerning the temperature data and the actual
warming, and the likely role of the Sun \citep{Connolly_2021,Scafetta_2023c,Scafetta_2023b,Soon2023}
and natural oscillations \citep{Scafetta_2013,Scafetta_2021} that
are not properly modeled by the GCMs. This article only evaluates
``realistic'' climate change projections while assuming valid the
economic and welfare models used by the IPCC \citep{IPCC_2022b} to
assess the impacts and risks that might be associated with projected
levels of global warming.

\section{Understanding the uncertainties}

Future climate-change impacts and risks are determined by the extent
to which the global surface will warm during the 21st century. Projected
climatic changes are determined by the amount of greenhouse gases
emitted and by how the climate system responds to them \citep{Hausfather2022}.
However, the anthropogenic signal should be viewed as superimposed
on top of natural climate variability, which consists of cycles and
fluctuations that occur at all timescales; the natural variability
occurring from the decadal to the millennial scales is especially
important for correctly interpreting the climatic changes observed
since 1850 \citep[e.g.:][]{Scafetta_2013,Scafetta_2021}.

\begin{figure}[!t]
\centering{}\includegraphics[width=1\textwidth]{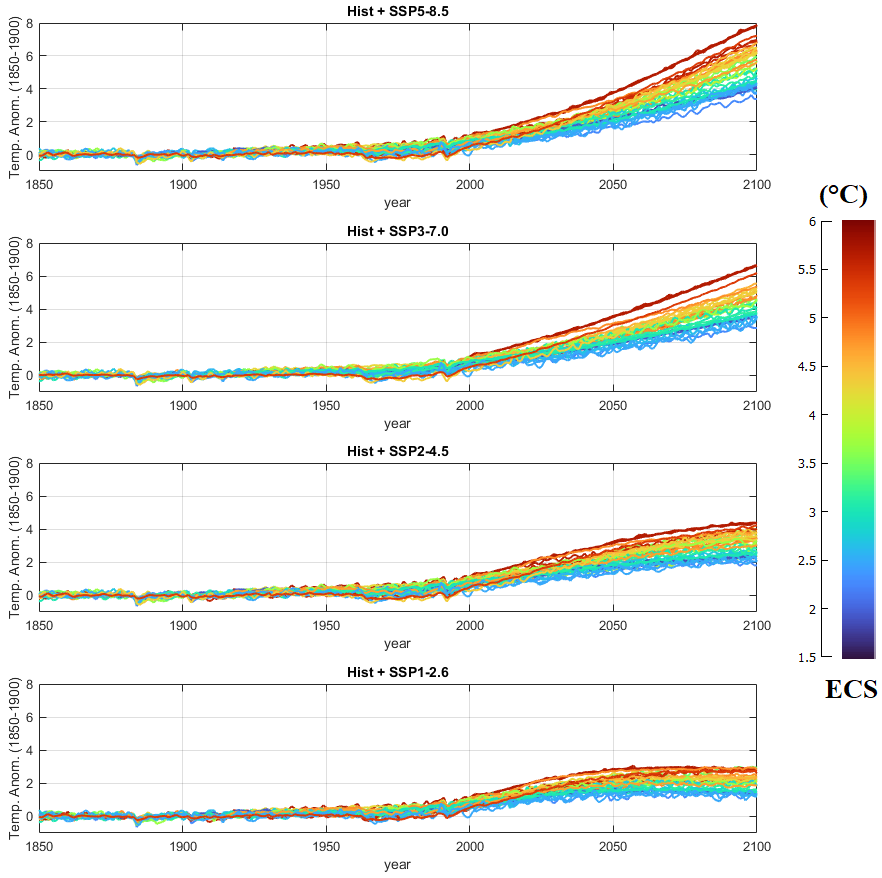}\caption{CMIP6 GCM simulations adopted in the present study (Appendix A). The
curve's color scales with the ECS of the models.}
\label{Fig1}
\end{figure}

GCM simulations for the 21st century, such as those shown in Figure
\ref{Fig1} \citep{IPCC_2021,IPCC_2022a}, were used to assess global
and regional impacts and risks associated with global warming. Figure
\ref{Fig1} illustrates that the GCM simulations differ significantly
both amongst models and because of the chosen SSP scenario adopted
for the simulation. The entire ensemble of climate projections estimates
that, by 2100, global surface temperatures might rise by 1.3 to 8.0
°C, meaning that climate change could have a wide range of effects
on civilizations and the environment throughout the 21st century,
ranging from beneficial to catastrophic. The main sources of uncertainty
are briefly described below.

\subsection{The SSP scenarios}

The GCM simulations for the 21st century were obtained under different
greenhouse gas emission scenarios known as representative concentration
pathways (RCPs) and, more recently, five shared socioeconomic pathways
(SSPs) describing alternative socioeconomic developments \citep{Riahi_2017}.
The latter were classified as SSP1 (sustainability), SSP2 (middle
of the road), SSP3 (regional rivalry), SSP4 (inequality), and SSP5
(fossil-fueled development). These labels were combined with the expected
level of radiative forcing in the year 2100 relative to 1750, which
varies from 1.9 to 8.5 W/m\textsuperscript{2}.

The SSP scenarios explicitly adopted by the IPCC AR6 \citep{IPCC_2021}
and herein considered are:
\begin{itemize}
\item SSP1-2.6 --- low GHG emissions and strong adaptation and mitigation
(CO\textsubscript{2} emissions are cut to net-zero around 2050--2075);
this is a kind of SSP scenarios compatible with the Paris Agreement
pledges needed to limit the global surface warming just below the
``safe'' threshold of 2 °C \citep{Meinshausen} by using the CMIP6
GCMs \citep{IPCC_2018,Hausfather_2022,Hausfather2022}.
\item SSP2-4.5 --- intermediate GHG emissions and moderate adaptation and
mitigation (CO\textsubscript{2} emission rates remain around current
levels until 2050 then they fall but do not reach net-zero by 2100);
\item SSP3-7.0 --- high GHG emissions and average no policy (CO\textsubscript{2}
emissions double by 2100);
\item SSP5-8.5 --- very high GHG emissions and worst-case no policy (CO\textsubscript{2}
emissions triple by 2075).
\end{itemize}
The likelihood of the aforementioned scenarios were not estimated
by the IPCC AR6 \citep[pages 54, 231 and 238]{IPCC_2021},\LyXHairSpace{}
despite the fact that having done so would have been essential for
properly assessing plausible climate-change impacts and risks. In
particular, the SSP5-8.5 scenario produces the most alarming global
warming projections ranging from 4 to 8 °C by 2100, and it has been
the most popular climate scenario used in the scientific literature
\citep{Pielke2021}. For example, the IPCC AR6 \citep[table 12.12]{IPCC_2021}
highlighted the assessments of the emergence of climate impact drivers
(CIDs) for the 21st century using the climate simulations derived
from the SSP5-8.5 scenario.

However, the choice of highlighting the climate simulations and their
risk assessments made with the SSP5-8.5 scenario is questionable because
a few authors have investigated the SSP likelihood issue and also
the IPCC AR6 briefly acknowledged that ``\emph{the likelihood of
high-emissions scenarios such as RCP8.5 or SSP5-8.5 is considered
low in light of recent developments in the energy sector}'' and that
``\emph{studies that consider possible future emissions trends in
the absence of additional climate policies, such as the recent IEA
2020 World Energy Outlook ‘stated policy’ scenario (IEA, 2020), project
approximately constant fossil fuel and industrial CO}\textsubscript{\emph{2}}\emph{
emissions out to 2070, approximately in line with the intermediate
RCP4.5, RCP6.0 and SSP2-4.5 scenarios}'' \citep[pages 238-239]{IPCC_2021}.
This is an important statement since the IPCC and many works have
considered the SSP5-8.5 extreme scenario to represent the ‘\emph{business
as usual}’ case \citep{Stocker} despite it was meant to only be a
very high-end, although still plausible, no-policy scenario \citep{vanVuuren}.

In fact, \citet{Ritchie2017} affirmed that the ``\emph{SSP5-RCP8.5
should not be a priority for future scientific research or a benchmark
for policy studies}'' because it is very unlikely because such scenario
is based on the implausible assumption of a massive increase of the
use of coal that is supposed to substitute all the other forms of
energy production. \citet{Hausfather_2020} stated that the\emph{
}``\emph{the ‘business as usual’ story is misleading}'' and invited
researchers ``\emph{to stop using the worst-case scenario for climate
warming as the most likely outcome}'' because only ``\emph{more-realistic
baselines make for better policy}''. \citet{Hausfather_2020} described
SSP5-8.5 as \emph{highly unlikely}, SSP3-7.0 as \emph{unlikely} (as
it requires the reversal of some current policies) and SSP2-4.5 as
\emph{likely} (because it is the most consistent with current policies).

The SSP5-8.5 and SSP3-7.0 scenarios were also recognized as unrealistic
by \citet{Burgess_2020} and \citet{Pielke_2021a,Pielke_2021b}, who
argued that the IPCC might have misused these scenarios for more than
a decade. \citet{Pielke_2022} and \citet{Burgess} also found that
another and even more moderate SSP scenario --- the SSP2-3.4, which
was ignored by the IPCC AR6 --- optimally agrees with the reported
emissions from 2005 to 2020 and with the most recent projections of
the \citet{IEA_2020,IEA_2021} up to 2050. This would imply that the
world could be on pace for even more moderate global warming by 2100
than what SSP2-4.5 implies.

For example, SSP3-7.0 and SSP5-8.5 demand that CO\textsubscript{2}
atmospheric concentrations climb from around 420 ppm to almost 867
ppm and 1135 ppm, respectively, from 2020 to 2100, which looks to
be excessively high; yet, SSP2-4.5 requires a more plausible 602 ppm
level by 2100 \citep{Meinshausen2020}. See also \citet{Meinshausen2023}
who reviewed the next generation of Earth system model likely scenarios.

As a result, only the SSP2-4.5 baseline or other even more moderate
scenarios \citep{Pielke_2022} could be plausible. Thus, policies
aimed to address realistic impacts and risks of climate changes should
be based only on such moderate scenarios.

\subsection{The GCM climate sensitivity uncertainty}

The latest computer climate simulations were generated using the Global
Climate Models (GCMs) of the Coupled Model Intercomparison Project
phase 6 (CMIP6), which is overseen by the World Climate Research Program
(WCRP) of the Working Group on Coupled Modelling (WGCM) \citep{Eyring_2016}.
More than 50 CMIP6 models were adopted to project future climates.
The models used historically estimated forcing functions from 1850
to 2014, which were extended from 2015 to 2100 using the estimated
forcing functions deduced from the chosen SSP scenarios. The initial
conditions and other internal model parameters could be randomly varied
within their estimated error ranges to produce ensembles of climate
simulations. The impacts and risks of the projected climate changes
are typically evaluated for prospective reference periods in the short
(2021--2040), mid-term (2041--2060), and long-term (2081--2100).
The ensembles of such climate projections were then used for policy
planning on adaptation, mitigation, and resilience.

However, as Figure \ref{Fig1} shows, the CMIP6 GCM simulations produce
diverging climate chronologies for the 21st century despite using
equivalent forcing functions. This significantly raises the level
of uncertainty in calculating the impacts and risks of future climate
change; such uncertainty needs to be restricted given the significant
societal costs associated with climate policies.

A core issue is that a significant fraction of GCMs appear to be functioning
too ``\emph{hot}'' \citep{Hausfather_2022}, implying that climate
change policy should discount climate-change projections from the
too-hot models. This problem was also acknowledged in the IPCC AR6
reports where the projections from the evidently too-hot GCMs were
not used for impact and risk assessments \citep{IPCC_2022a}. In fact,
the CMIP6 GCMs suggest that doubling atmospheric CO\textsubscript{2}
concentrations from pre-industrial levels (from 280 to 560 ppm) would
cause, at equilibrium, a global surface warming ranging from 1.8 to
5.7 °C \citep{IPCC_2021}. ECS values larger than 5 °C, on the other
hand, already exceeded those of the previous versions of global climate
models (CMIP3 and CMIP5 GCMs) used in the IPCC reports in 2007 and
2013.

Doubling the CO\textsubscript{2} content in the atmosphere results
in an increase in radiative forcing of around 3.7 W/m\textsuperscript{2},
which should cause a global warming of roughly 1 °C according to the
Stefan--Boltzmann law \citep{Rahmstorf_2008}. However, the actual
warming depends also on the feedback response of the climate system
to the applied forcing \citep{Roe_2009}, but the magnitude of such
a response is still highly uncertain \citep{Knutti_2017}.

There are different definitions of climate sensitivity. The equilibrium
climate sensitivity (ECS) is the long-term temperature rise (equilibrium
global mean near-surface air temperature) that is expected to result
from a doubling of the atmospheric CO\textsubscript{2} concentration
\citep{IPCC_2021}. Once the CO\textsubscript{2} concentration has
stopped rising and the majority of the feedbacks have had time to
fully take effect, the model predicts a new global mean near-surface
air temperature level. However, after CO\textsubscript{2} has doubled,
it can take centuries or even millennia to reach a new equilibrium
temperature. Another commonly used definition is the transient climate
response (TCR), which is defined as the change in the global mean
surface temperature, averaged over a 20-year period, centered at the
time of atmospheric CO\textsubscript{2} doubling, in a climate model
simulation in which the atmospheric CO\textsubscript{2} concentration
increases at 1\% per year until it doubles from 280 to 560 ppm \citep{IPCC_2021}.
This means to force the computer climate model with a linearly increasing
forcing from 0 to 3.7 $W/m^{2}$ for 70 years. As a result, TCR can
be assessed using simulations that last a shorter time than those
needed to assess ECS. In addition, TCR is smaller than ECS also because
slower feedbacks, which might further enhance a temperature increase,
require more time to fully respond to an increase in radiative forcing.
For instance, after a radiative perturbation, the deep ocean could
take several centuries to achieve a new steady state while continuing
to cool the top ocean as a heatsink. The CMIP6 GCMs calculate TCR
values ranging between 1.2 and 2.8 °C. In any case, ECS and TCR values
of the CMIP6 GCMs are closely correlated \citep{Scafetta_2023}.

Estimating the most realistic ECS and TCR ranges, on the other hand,
has been a highly controversial topic for more than 140 years because
their values strongly dependent on the physical assumptions that one
makes, particularly on how the water vapor and cloud feedbacks are
treated. For example, \citet{Arrhenius_1896} estimated that doubling
CO\textsubscript{2} could result in a 5-6 °C increase in global surface
temperature; however, 10 years later, the same author concluded that
his previous estimate was incorrect and proposed lower ECS values
ranging from 1.6 to 3.9 °C \citep{Arrhenius_1906}. \citet{Moller_1963}
found that the ECS estimate could vary greatly, up to one order of
magnitude, according to how water vapor and/or cloudiness responded
to the CO\textsubscript{2} perturbation, and concluded that such
an uncertainty implied that ``\emph{the theory that climatic variations
are affected by variations in the CO}\textsubscript{\emph{2}}\emph{
content becomes very questionable}''. \citet{Manabe_1967} developed
a one-dimensional climate model and estimated that the ECS could be
about 2.0 °C (which was related also to the 2.0 °C target for cost-benefit
analyses; \citealp{Gao}). Later, however, with a different model,
\citet{Manabe_1975} estimated that the likely ECS could be 2.93 °C.

In 1979, the US National Research Council concluded that ECS could
\emph{likely} (or 66\%) lie from 1.5 and 4.5 °C \citep{Charney1979}.
In its two earlier Assessment Reports --- FAR (1990) and SAR (1996)
--- the IPCC concurred and qualitatively estimated that the real
ECS value had likely to lie between 1.5 and 4.5 °C. In its third Assessment
Report, the IPCC changed a little bit the likely range to lie between
1.75 and 4.25 °C. In 2007, the IPCC Fourth Assessment Report claimed
to have formally quantified the uncertainties and that the likely
ECS range was to lie between 2 and 4.5 °C. In 2013, however, the IPCC
Fifth Assessment Report claimed again that the ECS likely range had
to be between 1.5 and 4.5 °C. By accepting the analysis by \citet{Sherwood_2020}
on the assessment of the Earth's climate sensitivity to radiative
forcing using multiple lines of evidence, the latest Assessment Report
of the IPCC \citep[AR6,][]{IPCC_2021} concluded that the likely ECS
(66\% confidence) should lie between 2.6 and 3.9\LyXThinSpace °C,
and very likely (90\% confidence) between 2.3 and 4.7\LyXThinSpace °C.
The likely TCR could range between 1.4 and 2.2 °C. Thus, the IPCC
AR6 concluded that the ECS \emph{likely} range had to lie between
2.5 and 4.0 °C, and \emph{very-likely} between 2.0 and 5.0 °C. This
suggested that to assess climate-related impacts and risks for the
21st century and, as a result, to develop public policies aimed at
addressing and/or mitigating them, the CMIP6 GCMs that predicted ECS
or TCR values outside of such ranges do not have to be used. \citet{Rugenstein}
provided an overview of the ``best'' ECS estimations proposed since
1979 to the present, all of which offer extremely broad uncertainty
ranges.

However, as additional studies were concluded, in time the task of
determining the actual ECS has grown more and more compelling. In
fact, \citet{Knutti_2017} stressed that not only the scientific literature
proposes ECS values ranging from 0.5 to 6.0 °C or even within a wider
range, but that ``\emph{evidence from climate modelling favours values
of ECS in the upper part of the 'likely' range, whereas many recent
studies based on instrumentally recorded warming --- and some from
palaeoclimate --- favour values in the lower part of the range}''.
\citet{Rugenstein} confirmed the conundrum by stating: ``\emph{Early
in the 2010s, a substantial discrepancy was noted between estimates
of climate sensitivity derived from climate models and estimates based
on the observed warming record and radiative balance ... Estimates
based on observed warming pointed to much lower values than those
derived from models}''. Such conclusions indicate that there is a
substantial dichotomy between empirical and GCM studies regarding
the actual climate sensitivity to radiative forcing, which is likely
owing to the GCMs' inability to adequately hindcast natural climate
variability and its geographic patterns. The dilemma is serious since,
according to the scientific method, conclusions based on empirical
data should be given preference, but the IPCC and climate policies
rely on impact and risk assessments derived from questionable GCM
projections. For example, \citet{Mauritsen} and \citet{Mauritsen(2020)}
demonstrated that by slightly altering the free parameters that regulate
the cloud feedback, the model ECS could be reduced from 7 °C to roughly
3 °C, effectively halving the expected global warming by 2100.

Several recent attempts have been made to better constrain the ECS
likely range. For example, \citet{Nijsse_2020} derived that the likely
ECS range should be 1.9-3.4 °C and \citet{Lewis_2023} directly challenged
the assessments proposed by \citet{Sherwood_2020}, which were adopted
by the IPCC \citep{IPCC_2021}. Lewis’ corrections and revisions of
the methodologies used by \citet{Sherwood_2020} lead to a likely
(17--83\%) ECS range between 1.75 and 2.7 °C and a 5--95\% range
between 1.55 and 3.2 °C. \citet{Scafetta_Climate_2021,Scafetta_GRL_2022,Scafetta_CliDyn_2022,Scafetta_2023}
tested how well the CMIP6 GCMs hindcast the global surface warming
observed from 1980--1990 to 2011--2021 in function of their ECS
and TCR values. Such a period was chosen because it is claimed to
be characterized by climate records with the smallest uncertainty
and, moreover, both surface and satellite records are available for
comparison and alternative interpretations. It was discovered that,
as climate sensitivity to radiative forcing diminishes, the performance
of the CMIP6 GCMs improves. By aggregating the CMIP6 GCMs into three
sub-ensembles (herein labeled ``macro-GCMs'') according to their
ECS value --- low-ECS ($1.5<\mathrm{ECS}\leq3.0$ °C), medium-ECS
($3.0<\mathrm{ECS}\leq4.5$ °C), and high-ECS ($4.5<\mathrm{ECS}\leq6.0$
°C) --- only the low-ECS macro-GCM was found to produce a warming
of $0.60\pm0.12$ °C, which hindcasts sufficiently well the 0.5-0.6
°C warming observed in various global surface temperature records
\citep{Hersbach_2020,Ishihara_2006,Lenssen_2019,Morice2012,Morice_2021,Rohde_2020,Zhang_2019}.
On the contrary, the medium and high-ECS macro-GCMs produced a warming
of $0.79\pm0.10$ °C and $0.94\pm0.23$ °C, respectively, which consistently
showed a warm bias, suggesting that these two macro-GCMs may not be
reliable climate predictors for climate-change policies for the 21st
century. Also \citet{Spencer2023} found that the actual value of
the ECS should be on the lower end of the range produced by current
GCMs and, therefore, confirmed \citet{Scafetta_GRL_2022}. It is worth
noting that \citet{Schmidt2023}'s critique of \citet{Scafetta_GRL_2022}
was refuted because it was shown to be based on statistical and physical
flaws \citep{Scafetta2023d}.

Thus, current research suggests that only the low-ECS macro-GCM could
be used for assessing the impacts and risks of climate change for
the 21st century because the actual ECS should be lower than about
3 °C while the IPCC AR6 assumed an ECS likely range between 2.5 and
4.0 °C.

\subsection{The surface and satellite 1980--2022 global warming divergence}

\citet{Scafetta_CliDyn_2022,Scafetta_2023} also argued that even
the low-ECS macro-GCM should be considered \emph{too hot} if the 0.4
°C of warming reported by the UAH-MSU satellite lower troposphere
(lt) global temperature record \citep{Spencer_2017} from 1980--1990
to 2011--2021 accurately represents actual surface warming better
than what global surface temperature records claim. In fact, \citet{McKitrick_2020}
and \citet{Mitchell_2020} discovered that all CMIP6 GCMs appear to
be affected by a strong warm bias throughout the troposphere, particularly
above the tropics and after 2000, where the models hindcast a hotspot
at roughly 10 km height that is not observed in the data. The difference
in model and satellite tropospheric warming rates was acknowledged
by \citet{Santer} and is reported by the IPCC AR6 \citep[figure 3.10]{IPCC_2021}.
The logical conclusion would be that the actual ECS value could be
at or below the bottom of the CMIP6 GCM range, that is below about
2 °C, which would imply an even lower projected warming for the 21st
century.

In general, according to the atmospheric physics used in the models,
the troposphere should warm quicker than the surface since greenhouse
gases are expected to warm the troposphere first by activating water
vapor and negative lapse rate feedbacks. As a result, the warming
of the lower troposphere should be larger, not smaller, than the warming
at the surface. Thus, if the surface warming of the Earth were caused
by greater greenhouse gas concentrations, as the GCMs suggest, the
warming rate in the lower troposphere may be viewed as the upper limit
of the possible warming rate at the surface.

The discrepancy between satellite-based and surface-based surface
temperature data could be due to non-climatic warm biases generated
by contamination from urban heat island (UHI) or other local non-climatic
warming sources existing on the surface. In fact, \citet{Scafetta_Ouyang_2019}
and \citet{Scafetta_CliDyn_2021} demonstrated that non-climatic warming
biases could affect extended land regions because, in comparison to
GCM simulations, nocturnal minimum temperatures warmed too quickly
relative to diurnal maximum temperatures, even in already homogenized
temperature records. Furthermore, comparisons of surface temperature
records between GCM simulations and lower troposphere UAH-MSU satellite
measurements revealed that the land region warmed far more quickly
than the ocean region. By contrast, the warming of the ocean area
is roughly replicated by the satellite microwave derived temperatures
\citep{Scafetta_CliDyn_2021,Scafetta_CliDyn_2022,Scafetta_2023}.
\citet{Scafetta_Climate_2021} also investigated and compared alternative
surface temperature data (ERA5-T2m, HadCRUT4, GISTEMP 250km, NOAA
v5) utilizing regional distributions of warming patterns from 1980--1990
to 2011--2021; it was discovered that the land regions were frequently
much warmer than what the UAH-MSU v. 6.0 lt and the ERA5 t850 temperature
records revealed. There exists also a so-called ``\emph{divergence
problem}'', in which proxy temperature records based on three-ring
width chronologies show much less warming than the temperature data
from land stations of similar latitudes since the 1970s \citep{Buntgen_2021,Spencer_2017},
which further suggests a non-climatic warm bias affecting global surface
temperature records \citep{Buntgen_2021,Spencer_2017}. Indeed, \citet{Connolly_2021}
and \citet{Soon2023} showed that a Northern Hemisphere land surface
temperature record derived entirely from confirmed rural-only stations
demonstrates up to 40\% less warming relative to the preindustrial
period (1850--1900) than records derived from both urban and rural
stations. Additional studied on a significant UHI influence on the
surface temperature records are being conducted \citep{Spencer(2023)}.

\citet[figure 5F]{Scafetta_2023} also compared three consecutive
versions of the HadCRUT global surface temperature records --- HadCRUT3
(discontinued after May 2014), HadCRUT4 (discontinued after December
2021), and non-infilled data and infilled HadCRUT5 (the most recent
version) --- and discovered that the global surface temperature warming
trend since 2000 has been progressively increased by consecutive adjustments
of the data. The 2000--2014 temperature ``\emph{pause}'' or ``\emph{hiatus}'',
which was clearly visible in HadCRUT3 and acknowledged by the IPCC
AR5 \citep{Stocker}, has now completely disappeared in the infilled
HadCRUT5 record, which adopts model predictions for filling the regional
and temporal gaps in the observations. This situation cannot be explained
simply as the result of optimized temperature data because the lower
troposphere satellite microwave derived temperature records (UAH-MSU
v6) still indicate a significant ``pause'' from 2000 to 2014. The
same is true for \citet{Connolly_2021} and \citet{Soon2023}'s rural-only
derived surface temperature record: see \citet[figure 5B]{Scafetta_2023}.
It is possible that the new adopted homogenization algorithms added
some spurious warming to the land records as a result of a mathematical
artifact known as ``\emph{urban blending}'' \citep{Katata,O=002019Neill},
and/or the models used to fill in the temporal and areal data gaps
introduced some spurious warming because of the warming bias that
the climate models present.

It should, however, be noticed that the satellite-based UAH-MSU lt
temperature record by \citet{Spencer_2017} was deemed controversial
because, after some data adjustments made in 2014 in the remote sensing
system (RSS), the RSS alternative lower troposphere global temperature
record by \citet{Mears_2016} showed a warming trend that appeared
more compatible with those presented by the surface-based temperature
records; the NOAA-STAR v. 4.0 troposphere temperature dataset presented
trends similar to RSS \citep{Santer} and, therefore, yielded a similar
conclusion contradicting UAH-MSU. However, \citet{Zou_2023} recently
revised the NOAA-STAR troposphere temperature records by addressing
certain errors in their previous version. The new NOAA-STAR v. 5.0
troposphere records now closely reflect and, therefore, confirms the
lower warming trend of the UAH-MSU temperature records. The discovery
adds to the growing body of evidence that global surface temperature
data are warm-biased, as the CMIP6 GCMs predict that the troposphere
should warm faster than the surface.

\subsection{Issues related to the natural climate variability not reproduced
by the GCMs}

Climate records are characterized by large secular and millennial
oscillations throughout the Holocene \citep[e.g.:][]{Alley,Bond_2001,Kutschera,Neff},
with the millennial ones responsible for well documented warm periods
such as the Roman (approximately 2000 years ago) and Medieval (around
1000 years ago) ones \citep{Ljungqvist2010,Luterbacher2016,Christiansen2012,Lasher,Buntgen_2021}.
Moreover, spectral examinations of global (land and ocean) surface
temperature data show that the climate system is characterized by
a significant number of distinct natural oscillations (for example,
with periods of roughly 5.2, 5.9, 6.6, 7.4, 9.1, 10.5, 14, 20 e 60
years) \citep{Scafetta2010,Scafetta_2013,Scafetta_2021}. The quasi
60-year oscillation appears to be rather important and is found in
a vast range of climatic datasets such as global sea level, Atlantic
Multidecadal Oscillation (AMO), and North Atlantic Oscillation (NAO)
records dating back for centuries and millennia \citep{Knudsen,Scafetta2014a,Wyatt_2013}.
The Holocene also experienced a Climate Optimum, particularly in the
Northern Hemisphere, between roughly 9,500 and 5,500 years ago, with
a thermal maximum around 8000 years ago, which is not reproduced by
models, indicating the need to improve our understanding of natural
climate forcings and feedbacks, as well as their representation in
GCMs \citep{Kaufman}.

The quasi millennial and 60-year oscillations appear to have been
responsible for at least 50\% of the 1900--2000 and 1970--2000 warming,
respectively. The latter was also responsible for the substantial
warm period that occurred in the 1940s and subsequent cooling observed
globally from the 1940s to the 1970s \citep{Scafetta2010,Scafetta_2013}.
It has been observed that a synchronous quasi-60-year modulation appears
in specific solar activity reconstructions over the last 150 years
that closely match the quasi-60-year modulation observed in climate
data \citep{Connolly2023,Scafetta_2023c,Soon2023}. In general, all
climatic oscillations from the inter-annual to the multi-millennial
timescales appear to be spectrally coherent with solar and/or astronomical
oscillations \citep[and many others]{Czymzik,Eddy1976,Kerr_2001,Kirkby_2007,Neff,Scafetta2010,Scafetta2014b,Scafetta2020,Scafetta_2023b,Schmutz_2021,Steinhilber_2012}.

\citet{Scafetta2012,Scafetta_2013,Scafetta_2021} demonstrated that
the GCMs are unable to reproduce these natural oscillations and, as
a result, they incorrectly attribute nearly 100\% of the global surface
warming observed from 1850--1900 to the present and, in particular,
that observed from the 1970s to the 2000s, to anthropogenic greenhouse
emissions. Indeed, it appears that the millennial natural oscillation
has been warming the climate system since 1700, that is since the
coldest era of the Little Ice Age, and may have been responsible for
a large percentage of the global surface warming of the twentieth
century.

For example, Figure 3.2 of the IPCC AR6 \citep[pp. 432]{IPCC_2021}
highlights the inability of the CMIP6 climate models to hindcast the
Medieval Warm Period, which normally lasted from roughly 800 to 1350.
Thus, these models are not able to reproduce the millennial oscillation.
In fact, the IPCC AR6 (pp. 433) explicitly acknowledges that ``\emph{before
the year 1300}'' there are ``\emph{larger disagreements between
models and temperature reconstructions}'' even by using the paleoclimate
reconstruction proposed by the \citet{PAGES2k}, which appears to
attenuate the Medieval Warming in comparison to other paleotemperature
reconstructions \citep[e.g.:][]{Moberg,Loehle,Mann,Ljungqvist2010,Christiansen2012,Ge,Kutschera,Luterbacher2016,Luning2019}
as those reported by the IPCC AR5 \citep[figure 5.7a]{Stocker}. The
IPCC AR6 is ambiguous on the physical cause of such disagreements
because it only states (pp. 433) that they could just\emph{ ``be
expected because forcing and temperature reconstructions are increasingly
uncertain further back in time (specific causes have not yet been
identified conclusively)}''. More specifically, AR6 did not acknowledge
that several empirical studies have already suggested that the cause
could have been a Medieval high solar activity \citep{Bond_2001,Eddy1976,Kerr_2001,Kirkby_2007,Steinhilber_2012,Scafetta_2013,Scafetta_2023b},
whose real effect on the climate could not be reproduced by the CMIP6
models.

Several authors proposed empirical models for global climate change
based on the above results. Here I will review the main model that
I proposed because it passed several tests \citep{Scafetta2010,Scafetta_2013,Scafetta_Climate_2021,Scafetta_2023c}.
The empirical modeling of natural climate oscillations not reproduced
by the GCMs suggests that the climate system is significantly less
sensitive to variations in greenhouse gas concentrations while being
significantly oversensitive to variations in solar activity changes.
I estimated that the actual ECS could be between 1 and 2 °C. Equally
low ECS estimates were also found by others \citep{Bates,Lindzen,McKitrick_2020,Monckton,Stefani}.

The climate system could appear to be over-sensitivity to solar activity
changes because the Sun should impact climate change not only through
total solar irradiance (TSI) variations, as the GCMs assume, but also
through variations in its ultraviolet light and magnetic field. UV
forcing directly modifies stratospheric ozone, whereas solar magnetic
activity modifies fluxes of interplanetary charged particles -- cosmic
rays, solar wind, and interplanetary dust -- that could directly
influence the electric properties of the atmosphere and its cloud
system \citep{ScafettaOrtolani,ScafettaMilani,Scafetta_2023c,Shaviv_2002,Svensmark_2016,Svensmark_2022}.
In particular, \citet{Scafetta_2023c} showed that direct TSI forcing
could even explain only 20\% of the total solar effect on climate
change. That TSI variations by alone cannot explain how to Sun could
modulate the atmosphere of a planet was recently demonstrated by \citet{Chavez},
who discovered that, despite Neptune being 30 times farther from the
Sun than Earth, its cloud activity exhibits remarkable relationships
with the 11-year solar activity cycles; the (still unknown) physical
mechanism could not be TSI variations since on Neptune the latter
are 900 times smaller than on Earth.

Furthermore, it should be noticed that the CMIP6 GCMs minimize the
solar effect on the climate also because they use a solar forcing
derived from a specific TSI reconstruction which presents a very little
and nearly constant secular variability \citep{Matthes_2017}, although
the scientific literature also recommends other TSI reconstructions
that show a much larger secular variability \citep[e.g.][]{Hoyt_1993,Egorova_2018,Penza_2022}
and that also appear to be better correlated with uncontroversial
TSI satellite measurements \citep{Scafetta_2019,Connolly_2021,Scafetta_2023c}.
Also the CMIP6 models used for the millennial simulation were forced
with a very low variability TSI reconstruction such as the PMIP4 SATIRE-M
\citep{Wu2018}, which could be erroneous because several other TSI
reconstructions exist and show a larger secular and millennial variability
\citep[cf.][]{Scafetta_2019,Scafetta_2023c,Connolly2023}. In particular,
\citet[figure 7]{Scafetta_2021} explicitly demonstrated that the
GCMs cannot reproduce the Medieval Warm Period (and likely any other
warm periods of the Holocene) because they assume only a nearly constant
radiative solar forcing and, therefore, there is no net radiative
forcing capable of reproducing a significant warming during the Medieval
period because volcanic forcing is sporadic and anthropogenic forcing
is also absent. This (apparently arbitrary) choice taken by the CMIP6
GCM modelers regarding solar forcing also contributes to both the
GCMs' high ECS values (which are required to reconstruct the 20th-century
warming using anthropogenic forcing) and to their failure in reconstructing
the other warm periods of the Holocene.

See \citet{Connolly_2021}, \citet{Scafetta_2023c} and \citet{Soon2023}
for additional commentary, modeling and supporting bibliography regarding
how to empirically assess of the role of the Sun in climate change
using various multi-proxy solar records.

\section{Addressing the \textquotedblleft too hot\textquotedblright{} model
problem}

As discussed in Section 2, a number of evidence point to the conclusion
that at least the majority of CMIP6 GCMs are running ``too hot'',
even more than what the IPCC \citep{IPCC_2021} and \citet{Hausfather_2022}
have already admitted. These findings suggest that the CMIP6 GCM climate
sensitivity to radiative forcings is significantly overestimated.
The CMIP3 and CMIP5 GCMs were already running too hot \citep{Scafetta2012,Scafetta_2013},
but the situation has gotten so critical with the CMIP6 GCMs that
also the UN climate panel has raised concerns \citep{Voosen2019,Voosen2021}.

The most recent research on direct comparison of the CMIP6 GCM hindcasts
with global surface temperature records since 1980 indicates that,
at most, only the CMIP6 GCMs with low climate sensitivity (e.g. for
ECS $\leq3$ °C) could adequately hindcast the observed warming since
1980 and could be used to evaluate the risk associated with projected
climate changes for the 21st century \citep{Scafetta_GRL_2022,Scafetta_CliDyn_2022,Scafetta2023d}.
Other authors have reached the same result \citep{Lewis_2023,Spencer2023}.
Furthermore, various evidences suggest that actual global warming
since 1980 may have been less than what is generally reported by the
global surface temperature records \citep{Scafetta_2023}. If this
is the case, even the low-ECS CMIP6 macro-GCM would be functioning
too hot, rendering all CMIP6 GCMs untrustworthy for guiding public
policy targeted at addressing the impacts and risks of projected climate
changes for the 21st century. Finally, alternative climate modeling
based on actual evidence of underestimated natural cycles and solar
influences also suggests low ECS values to radiative forcing. In fact,
according to these evidences, the true ECS may likely be in the range
of 1 to 2 °C \citep[e.g.:][]{Scafetta_2013,Scafetta_2021,Scafetta_2023c,Scafetta_2023b}.

In the following, I will assess the impacts and risks of climate change
suggested by the IPCC \citep{IPCC_2022a} using the SSP2.4.5 scenario
and the 21st-century warming projected by the low-ECS CMIP6 macro-GCM.
I also provide alternative calculations that account for the possibility
that even the low sensitivity models are running too hot because,
based on satellite low troposphere temperature records, the actual
1980--2022 warming may be less than what claimed in \citet{IPCC_2021}.
Finally, alternative risk assessments are provided by employing the
empirical global surface temperature models for the 21st century proposed
by \citet{Scafetta_2013,Scafetta_2021}.

\section{Re-assessment: selection of reliable GCMs}

Climate simulations from 42 CMIP6 GCMs are analyzed here. The computer
simulations were created using historical forcings (1850--2014) further
extended up to 2100 with hypothetical forcing functions deduced from
four different SSP scenarios: SSP1-2.6 (low GHG emissions), SSP2-4.5
(intermediate GHG emissions), SSP3-7.0 (high GHG emissions) and SSP5-8.5
(very high greenhouse gas emissions). These four scenarios are nearly
indistinguishable until 2022. Thus, from 1850 to 2022, the four simulation
ensembles can be considered independent assessments of the same models
under nearly identical forcing conditions. This variability also helps
to assess in first approximation the spreading related with the chaotic
internal variability of the models: see also \citet{Scafetta_CliDyn_2022}
where the GCM internal variability issue is extensively discussed.

A total of 156 simulations were analyzed. 142 simulations were “average”
records provided by the Koninklijk Nederlands Meteorologisch Instituut
(KNMI) Climate Explorer \citep{Oldenborgh_2020}. Other 14 simulations
were missing in Climate Explorer and were taken from the supplementary
of \citet{Hausfather_2022}. The latter include those produced by
four GCMs: CMCC-ESM2, GFDL-CM4, IITM-ESM and TaiESM1. The ECS and
TCR values of the GCMs were taken from Table 7.SM.5 of the IPCC AR6
\citep{IPCC_2021}. Some missing values were taken from the supplementary
of \citet{Hausfather_2022}. The ECS of FIO-ESM-2-0 was hypothesized
to be 4.27 °C because its TCR falls between those of CNRM-CM6-1 and
EC-Earth3.

\begin{table}[!t]
\centering{}{\small{}}%
\begin{tabular}{|l|lc||l|lc|}
\hline 
Macro-GCM & ﻿Model Name & ECS (°C) & Macro-GCM & Model Name & TCR (°C)\tabularnewline
\hline 
\multirow{15}{*}{High-ECS} & CIESM & 5.63 & \multirow{12}{*}{High-TCR} & UKESM1-0-LL & 2.79\tabularnewline
 & CanESM5 p1 & 5.62 &  & CanESM5 p1 & 2.74\tabularnewline
 & CanESM5 p2 & 5.62 &  & CanESM5 p2 & 2.74\tabularnewline
 & CanESM5-CanOE p2 & 5.62 &  & CanESM5-CanOE p2 & 2.74\tabularnewline
 & HadGEM3-GC31-LL-f3 & 5.55 &  & NESM3 & 2.72\tabularnewline
 & HadGEM3-GC31-MM f3 & 5.42 &  & EC-Earth3-Veg & 2.62\tabularnewline
 & UKESM1-0-LL & 5.34 &  & HadGEM3-GC31-MM f3 & 2.58\tabularnewline
 & CESM2 & 5.16 &  & HadGEM3-GC31-LL f3 & 2.55\tabularnewline
 & CNRM-CM6-1 f2 & 4.83 &  & CNRM-CM6-1-HR f2 & 2.48\tabularnewline
 & CNRM-ESM2-1 f2 & 4.76 &  & CIESM & 2.39\tabularnewline
 & CESM2-WACCM & 4.75 &  & TaiESM1 & 2.34\tabularnewline
 & KACE-1-0-G & 4.75 &  & IPSL-CM6A-LR & 2.32\tabularnewline
\cline{4-6} \cline{5-6} \cline{6-6} 
 & ACCESS-CM2 & 4.72 & \multirow{16}{*}{Medium-TCR} & EC-Earth3 & 2.30\tabularnewline
 & NESM3 & 4.72 &  & FIO-ESM-2-0 & 2.22\tabularnewline
 & IPSL-CM6A-LR & 4.56 &  & CNRM-CM6-1 f2 & 2.14\tabularnewline
\cline{1-3} \cline{2-3} \cline{3-3} 
\multirow{13}{*}{Medium-ECS} & EC-Earth3-Veg & 4.31 &  & ACCESS-CM2 & 2.10\tabularnewline
 & TaiESM1 & 4.31 &  & CMCC-CM2-SR5 & 2.09\tabularnewline
 & CNRM-CM6-1-HR f2 & 4.28 &  & AWI-CM-1-1-MR & 2.06\tabularnewline
 & FIO-ESM-2-0 & 4.27 &  & CESM2 & 2.06\tabularnewline
 & EC-Earth3 & 4.26 &  & KACE-1-0-G & 2.04\tabularnewline
 & GFDL-CM4 & 3.89 &  & GFDL-CM4 & 2.00\tabularnewline
 & ACCESS-ESM1-5 & 3.87 &  & CESM2-WACCM & 1.98\tabularnewline
 & MCM-UA-1-0 & 3.65 &  & ACCESS-ESM1-5 & 1.95\tabularnewline
 & CMCC-ESM2 & 3.58 &  & FGOALS-f3-L & 1.94\tabularnewline
 & CMCC-CM2-SR5 & 3.52 &  & MCM-UA-1-0 & 1.94\tabularnewline
 & AWI-CM-1-1-MR & 3.16 &  & CMCC-ESM2 & 1.92\tabularnewline
 & MRI-ESM2-0 & 3.15 &  & CNRM-ESM2-1-f2 & 1.86\tabularnewline
 & BCC-CSM2-MR & 3.04 &  & MPI-ESM1-2-LR & 1.84\tabularnewline
\hline 
\multirow{14}{*}{Low-ECS} & FGOALS-f3-L & 3.00 & \multirow{14}{*}{Low-TCR} & GISS-E2-1-G p3 & 1.80\tabularnewline
 & MPI-ESM1-2-LR & 3.00 &  & CAMS-CSM1-0 & 1.73\tabularnewline
 & MPI-ESM1-2-HR & 2.98 &  & BCC-CSM2-MR & 1.72\tabularnewline
 & FGOALS-g3 & 2.88 &  & IITM-ESM & 1.71\tabularnewline
 & GISS-E2-1-G p3 & 2.72 &  & MPI-ESM1-2-HR & 1.66\tabularnewline
 & MIROC-ES2L & 2.68 &  & MRI-ESM2-0 & 1.64\tabularnewline
 & GFDL-ESM4 & 2.65 &  & GFDL-ESM4 & 1.63\tabularnewline
 & MIROC6 & 2.61 &  & MIROC-ES2L & 1.55\tabularnewline
 & NorESM2-LM & 2.54 &  & MIROC6 & 1.55\tabularnewline
 & NorESM2-MM & 2.50 &  & FGOALS-g3 & 1.54\tabularnewline
 & IITM-ESM & 2.37 &  & NorESM2-LM & 1.48\tabularnewline
 & CAMS-CSM1-0 & 2.29 &  & INM-CM5-0 & 1.41\tabularnewline
 & INM-CM5-0 & 1.92 &  & INM-CM4-8 & 1.33\tabularnewline
 & INM-CM4-8 & 1.83 &  & NorESM2-MM & 1.33\tabularnewline
\hline 
\end{tabular}\caption{CMIP6 GCMs analyzed in the present study. They are ranked according
to their ECS (left) and TCR (right) values. (The ECS of FIO-ESM-2-0
is hypothesized).}
\label{Tab1}
\end{table}

Table \ref{Tab1} lists the adopted GCMs with their estimated ECS
and TCR values, which are approximately correlated to each other \citep{Scafetta_2023}.
The GCMs are also grouped in three macro-GCMs according to their ECS
value: low-ECS ($1.5<\mathrm{ECS}\leq3.0\;{^\circ}\mathrm{C}$); medium-ECS
($3.0<\mathrm{ECS}\leq4.5\;{^\circ}\mathrm{C}$), and high-ECS ($4.5<\mathrm{ECS}\leq6.0\;{^\circ}\mathrm{C}$),
as already proposed by \citet{Scafetta_2023,Scafetta_CliDyn_2022,Scafetta_GRL_2022}.

Figure \ref{Fig1} shows the adopted 156 CMIP6 GCM simulations which
are baselined in 1850--1900. The curve's color scales with the ECS
value of the models from blue (low sensitivity) to red (high sensitivity).
The figure shows that, as the ECS increases, the warming projected
by the models during the 21st century tends to increase as well. Similar
results can be obtained by coloring the curves in function of the
TCR value of their GCM.

\begin{figure}[!t]
\centering{}\includegraphics[width=1\textwidth]{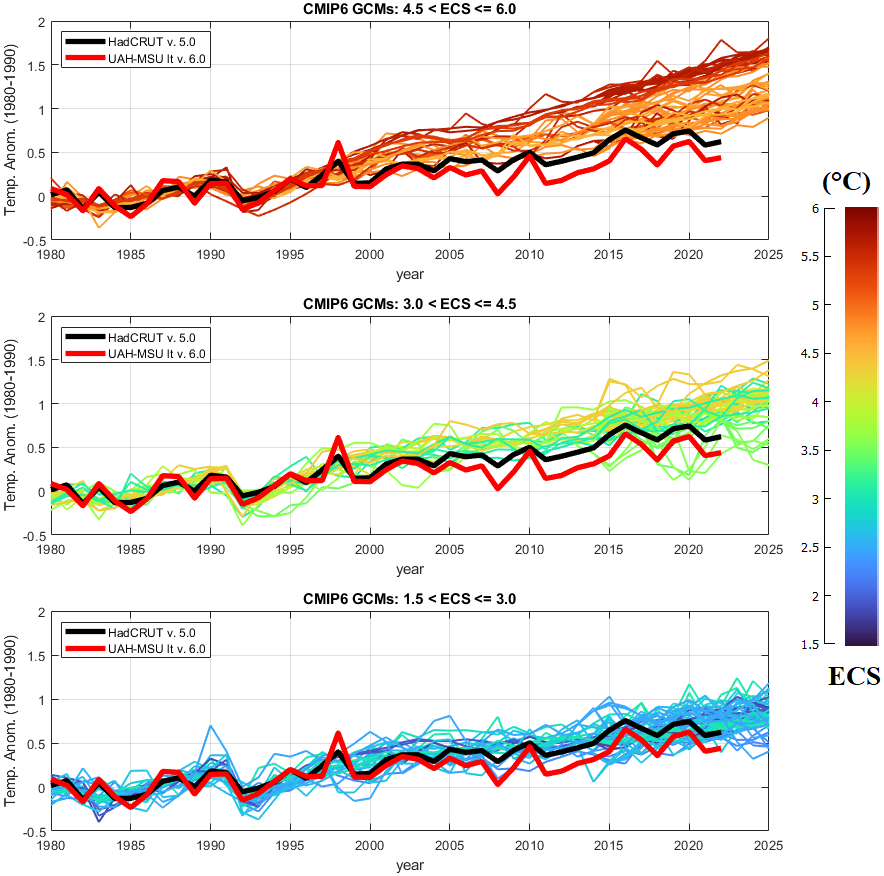}\caption{CMIP6 GCM simulations adopted in the present study divided into three
macro-GCMs according to their ECS value. The curve’s color scales
with the ECS of the models. The synthetic records are compared against
the HadCRUT v. 5.0 (infilled data) global surface temperature and
the satellite-based UAH-MSU lt v. 6.0 \citep{Spencer_2017}.}
\label{Fig2}
\end{figure}

Figure \ref{Fig2} shows the adopted 156 CMIP6 GCM simulations, which
are now baselined in 1980--1990 to better assess their performance
since 1980, when satellite temperature records are also available.
The data from the three macro-GCMs are displayed on the three panels.
The synthetic records are contrasted to the HadCRUT v. 5.0 global
surface temperature (infilled data, which present the most warming
of any other HadCRUT records) and the satellite-based UAH-MSU lt v.
6.0 \citep{Spencer_2017}, which grows warmer significantly less.

The warming from 1980--1990 to 2011--2022 shown by the (infilled)
HadCRUT5 record is nearly identical to that from other global surface
temperature records: ERA5-T2m \citep{Hersbach_2020}, GISTEMP \citep{Lenssen_2019},
and Berkeley Earth Land/Ocean temperature \citep{Rohde_2020}. Instead,
the warming presented by the UAH-MSU lower troposphere temperature
record is nearly identical to that shown by the recent NOAA-STAR v.
5.0 one \citep[data from][]{Zou_2023}.

Figure \ref{Fig2} reveals that, since 1980, simulations from the
medium and high-ECS macro-GCMs appear to be too hot. However, also
the low-ECS macro-GCM appears to produce too hot simulations in comparison
with the satellite UAH-MSU lt record. These results are confirmed
also by considering the 688 GCM individual ensemble simulations available
on Climate Explorer \citep[see][figures 1 and 2]{Scafetta_CliDyn_2022}.

\begin{figure}[!t]
\centering{}\includegraphics[width=1\textwidth]{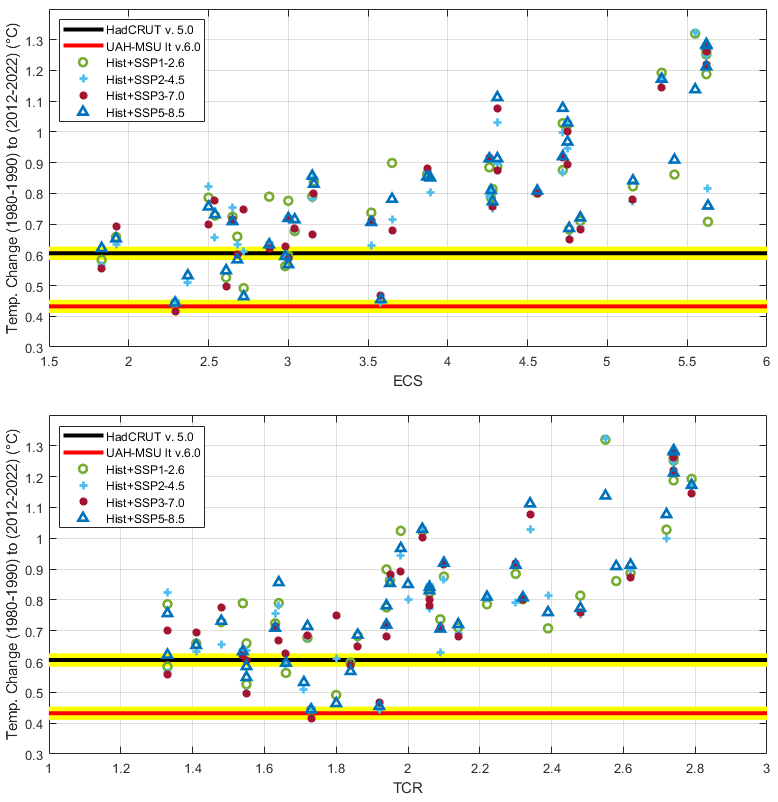}\caption{Temperature changes from 1980--1990 to 2012--2022 produced the adopted
42 GCMs against the warming of HadCRUT v. 5.0 ($0.605\pm0.02$ °C)
and UAH-MSU lt v. 6.0 ($0.432\pm0.03$ °C): see Table \ref{Tab2}.
Top: ECS ranking; Bottom: TCR ranking.}
\label{Fig3}
\end{figure}

Figure \ref{Fig3} confirms the above expectations by using the same
approach proposed by \citet{Scafetta_GRL_2022,Scafetta_CliDyn_2022,Scafetta_2023}.
It shows the temperature changes from 1980--1990 to 2012--2022 produced
by the adopted 42 GCMs against the correspondent warming shown by
HadCRUT v. 5.0 ($0.605\pm0.02$ °C, 95\% confidence) and UAH-MSU lt
v. 6.0 ($0.432\pm0.03$ °C, 95\% confidence). The top panel shows
the results against the ECS ranking; the bottom panel shows the same
against the TCR ranking. The figure demonstrates that, as ECS or TCR
increase, the warming hindcasted by the GCMs increases as well. The
GCM results are listed in Table \ref{Tab2}, which also reports the
{[}2.5\%, 17\%, 50\%, 83\%, 97.5\%{]} percentiles for each macro-GCM.

\begin{table}[!t]
\centering{}%
\begin{tabular}{|l|c|cccc|cc|}
\hline 
﻿ &  & \multicolumn{4}{c|}{Hist +} &  & \tabularnewline
Model & ECS (°C) & SSP1-2.6 & SSP2-4.5 & SSP3-7.0 & SSP5-8.5 & percentile & °C\tabularnewline
\hline 
CIESM & 5.63 & 0.71 & 0.81 & NaN & 0.76 & 97.5\% & 1.31\tabularnewline
CanESM5-p1 & 5.62 & 1.25 & 1.25 & 1.26 & 1.28 & 83.0\% & 1.24\tabularnewline
CanESM5-p2 & 5.62 & 1.26 & 1.28 & 1.28 & 1.29 & 50.0\% & 0.97\tabularnewline
CanESM5-CanOE-p2 & 5.62 & 1.19 & 1.21 & 1.22 & 1.21 & 17.0\% & 0.76\tabularnewline
HadGEM3-GC31-LL-f3 & 5.55 & 1.32 & 1.32 & NaN & 1.14 & 2.5\% & 0.66\tabularnewline
HadGEM3-GC31-MM-f3 & 5.42 & 0.86 & NaN & NaN & 0.91 &  & \tabularnewline
UKESM1-0-LL-f2 & 5.34 & 1.19 & 1.17 & 1.14 & 1.17 &  & \tabularnewline
CESM2 & 5.16 & 0.82 & 0.77 & 0.78 & 0.84 &  & \tabularnewline
CNRM-CM6-1-f2 & 4.83 & 0.71 & 0.69 & 0.68 & 0.72 &  & \tabularnewline
CNRM-ESM2-1-f2 & 4.76 & 0.68 & 0.65 & 0.65 & 0.69 &  & \tabularnewline
CESM2-WACCM & 4.75 & 1.02 & 0.95 & 0.89 & 0.97 &  & \tabularnewline
KACE-1-0-G & 4.75 & 1.02 & 1.03 & 1.00 & 1.03 &  & \tabularnewline
ACCESS-CM2 & 4.72 & 0.88 & 0.87 & 0.92 & 0.92 &  & \tabularnewline
NESM3 & 4.72 & 1.03 & 1.00 & NaN & 1.08 &  & \tabularnewline
IPSL-CM6A-LR & 4.56 & 0.80 & 0.80 & 0.80 & 0.81 &  & \tabularnewline
\hline 
EC-Earth3-Veg & 4.31 & 0.89 & 0.90 & 0.87 & 0.91 & 97.5\% & 1.07\tabularnewline
TaiESM1 & 4.31 & NaN & 1.03 & 1.08 & 1.11 & 83.0\% & 0.89\tabularnewline
CNRM-CM6-1-HR-f2 & 4.28 & 0.81 & 0.75 & 0.76 & 0.77 & 50.0\% & 0.80\tabularnewline
FIO-ESM-2-0 & 4.27 & 0.79 & 0.80 & NaN & 0.81 & 17.0\% & 0.68\tabularnewline
EC-Earth3 & 4.26 & 0.88 & 0.79 & 0.92 & 0.91 & 2.5\% & 0.46\tabularnewline
GFDL-CM4 & 3.89 & NaN & 0.80 & NaN & 0.85 &  & \tabularnewline
ACCESS-ESM1-5 & 3.87 & 0.86 & 0.87 & 0.88 & 0.85 &  & \tabularnewline
MCM-UA-1-0 & 3.65 & 0.90 & 0.71 & 0.68 & 0.78 &  & \tabularnewline
CMCC-ESM2 & 3.58 & NaN & 0.44 & 0.47 & 0.46 &  & \tabularnewline
CMCC-CM2-SR5 & 3.52 & 0.74 & 0.63 & 0.71 & 0.71 &  & \tabularnewline
AWI-CM-1-1-MR & 3.16 & 0.84 & 0.85 & 0.80 & 0.83 &  & \tabularnewline
MRI-ESM2-0 & 3.15 & 0.79 & 0.78 & 0.67 & 0.86 &  & \tabularnewline
BCC-CSM2-MR & 3.04 & 0.68 & 0.68 & 0.68 & 0.71 &  & \tabularnewline
\hline 
FGOALS-f3-L & 3.00 & 0.78 & 0.72 & 0.72 & 0.72 & 97.5\% & 0.79\tabularnewline
MPI-ESM1-2-LR & 3.00 & 0.60 & 0.59 & 0.59 & 0.57 & 83.0\% & 0.73\tabularnewline
MPI-ESM1-2-HR & 2.98 & 0.56 & 0.59 & 0.63 & 0.60 & 50.0\% & 0.62\tabularnewline
FGOALS-g3 & 2.88 & 0.79 & 0.64 & 0.62 & 0.63 & 17.0\% & 0.53\tabularnewline
GISS-E2-1-G-p3 & 2.72 & 0.49 & 0.61 & 0.75 & 0.46 & 2.5\% & 0.44\tabularnewline
MIROC-ES2L-f2 & 2.68 & 0.66 & 0.63 & 0.60 & 0.58 &  & \tabularnewline
GFDL-ESM4 & 2.65 & 0.72 & 0.76 & 0.72 & 0.71 &  & \tabularnewline
MIROC6 & 2.61 & 0.53 & 0.50 & 0.50 & 0.55 &  & \tabularnewline
NorESM2-LM & 2.54 & 0.73 & 0.66 & 0.78 & 0.73 &  & \tabularnewline
NorESM2-MM & 2.50 & 0.79 & 0.82 & 0.70 & 0.76 &  & \tabularnewline
IITM-ESM & 2.37 & NaN & 0.51 & NaN & 0.53 &  & \tabularnewline
CAMS-CSM1-0 & 2.29 & 0.44 & 0.45 & 0.41 & 0.44 &  & \tabularnewline
INM-CM5-0 & 1.92 & 0.66 & 0.63 & 0.69 & 0.65 &  & \tabularnewline
INM-CM4-8 & 1.83 & 0.58 & 0.56 & 0.56 & 0.62 &  & \tabularnewline
\hline 
\end{tabular}\caption{Temperature changes from 1980--1990 to 2012--2022 produced the adopted
42 GCMs, with percentiles for each macro-GCM. NaN indicates missing
simulations. (The ECS of FIO-ESM-2-0 is hypothesized).}
\label{Tab2}
\end{table}

Only the low-ECS macro-GCM ($1.5<ECS\lessapprox3.0$ °C and $1.3<TCR\lessapprox1.8$
°C) appears to generate a distribution of hindcasts that best incorporates
the observed warming. These ECS and TCR ranges significantly restrict
those of the CMIP6 GCMs (1.5-6.0 °C and 1.3-2.8 °C, respectively)
and are compatible with those estimated by \citet{Lewis_2023}. However,
if the actual global warming from 1980 to 2022 is compatible with
that shown by the lower troposphere estimates of UAH-MSU v. 6.0, Figure
\ref{Fig3} indicates that even the low-ECS macro-GCM would be running
too hot \citep[cf.][]{McKitrick_2020}. Each model is represented
by four separate simulations (where available), which provide a first
order estimate of the dispersion due to the models' internal variability.

Figure \ref{Fig3} also shows that a few GCMs with medium to high
ECS and TCR values may be approaching the observed warming. However,
because climate projections are created by multi-model ensembles rather
than individual models or simulations, a very tiny number of favorable
instances appearing in the distribution's tails should be statistically
classed as outliers. For example, CNRM-ESM2-1 appears to approach
the HadCRUT5 warming within the allowed range of $\pm0.1$ °C \citep[cf.][]{Scafetta_CliDyn_2022}
only because the model's extraordinarily high ECS (4.76 °C) is compensated
by its comparatively low TCR (1.86 °C), which slows down the global
surface temperature rise. However, relying on just one GCM to judge
whether such a scenario is conceivable would be insufficient; additional
GCMs must be created and statistically proven to be consistent with
both extremely high ECS values and relatively low TCR values.

\begin{figure}[!t]
\centering{}\includegraphics[width=1\textwidth]{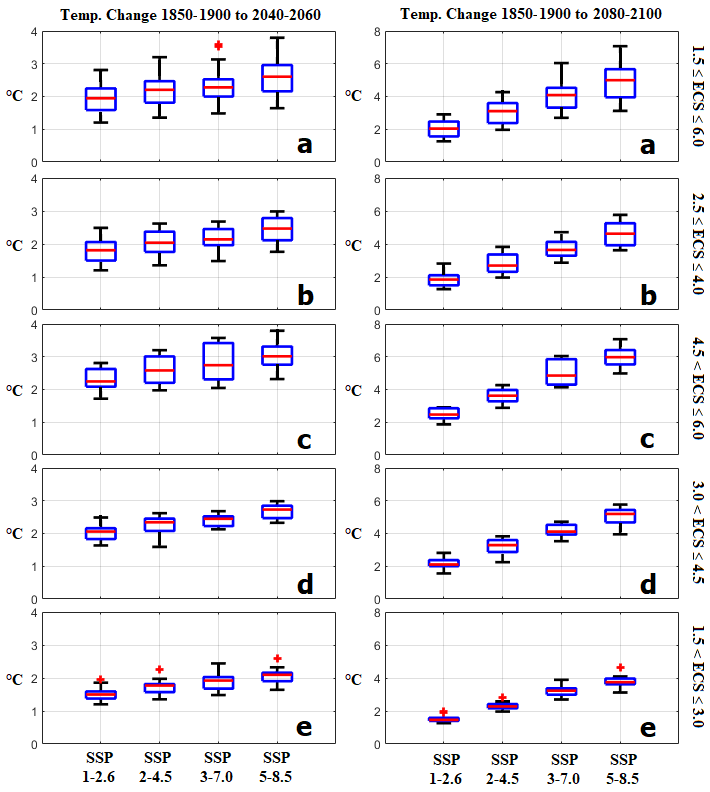}\caption{Boxplots indicating the temperature changes from 1850--1900 to 2040--2060
(left) and 2080--2100 (right) produced the adopted 42 GCMs divided
for the Hist + SSP1-2.6, SSP2-4.5, SSP3-7.0 and SSP5-8.5 emission
scenarios. (a) All CMIP6 GCMs; (b) CMIP6 GCMs with ECS ranging between
2.5 and 4.0 °C (the likely IPCC AR6 range); (c) High-ECS CMIP6 GCMs
($4.5<ECS\protect\leq6.0$ °C); (d) Medium-ECS CMIP6 GCMs ($3.0<ECS\protect\leq4.5$
°C); (e) Low-ECS CMIP6 GCMs ($1.5<ECS\protect\leq3.0$ °C).}
\label{Fig4}
\end{figure}

Figure \ref{Fig4} shows boxplots indicating the temperature changes
from 1850--1900 to 2040--2060 (left) and 2080--2100 (right) produced
the adopted 42 GCMs for the Hist + SSP1-2.6, SSP2-4.5, SSP3-7.0 and
SSP5-8.5 emission scenarios. The boxplots indicate the distribution
of the GCM forecasts based on a five-number summary (“minimum”, first
quartile {[}Q1 = 25\%{]}, median {[}Q2 = 50\%{]}, third quartile {[}Q3
= 75\%{]}, “maximum”) plus outliers, which are displayed as red crosses.
Each line of the figure represents a different selection of GCMs:
(a) is based on all CMIP6 GCMs; (b) uses only the GCMs with ECS ranging
between 2.5 and 4.0 °C, which corresponds to the IPCC AR6 likely range;
(c) uses only the high-ECS GCMs ($4.5<ECS\leq6.0$ °C); (d) uses only
the medium-ECS CMIP6 GCMs ($3.0<ECS\leq4.5$ °C); (e) uses only the
low-ECS CMIP6 GCMs ($1.5<ECS\leq3.0$ °C). Table \ref{Tab3} summaries
the quartile statistics of the temperature changes from 1850--1900
to 2040--2060 and 2080--2100 produced by the adopted 42 GCMs, similar
to that produced by the boxplots depicted in Figure \ref{Fig4} but
with a different percentage descriptions {[}2.5\%, 17\%, 50\%, 83\%,
97.5\%{]}, which highlights the median, the (likely) 66\% and the
(very likely) 95\% confidence ranges.

\begin{table}[!t]
\centering{}{\small{}}%
\begin{tabular}{|l|c|cc|cc|cc|cc|}
\hline 
\multicolumn{1}{|l}{{\small{}﻿}} &  & \multicolumn{2}{c|}{{\small{}Hist+SSP1-2.6 (}°C{\small{})}} & \multicolumn{2}{c|}{{\small{}Hist+SSP2-4.5 (}°C{\small{})}} & \multicolumn{2}{c|}{{\small{}Hist+SSP3-7.0 (}°C{\small{})}} & \multicolumn{2}{c|}{{\small{}Hist+SSP5-8.5 (}°C{\small{})}}\tabularnewline
\multicolumn{1}{|l}{} & {\small{}percentile} & {\small{}2040-2060} & {\small{}2080-2100} & {\small{}2040-2060} & {\small{}2080-2100} & {\small{}2040-2060} & {\small{}2080-2100} & {\small{}2040-2060} & {\small{}2080-2100}\tabularnewline
\hline 
\multirow{5}{*}{{\small{}Fig. \ref{Fig4}A}} & {\small{}97.5\%} & {\small{}﻿2.81} & {\small{}2.90} & {\small{}3.19} & {\small{}4.25} & {\small{}3.57} & {\small{}6.03} & {\small{}3.79} & {\small{}7.06}\tabularnewline
 & {\small{}83\%} & {\small{}2.40} & {\small{}2.76} & {\small{}2.62} & {\small{}3.72} & {\small{}2.71} & {\small{}4.79} & {\small{}3.03} & {\small{}5.98}\tabularnewline
 & {\small{}50\%} & {\small{}1.95} & {\small{}2.04} & {\small{}2.20} & {\small{}3.11} & {\small{}2.28} & {\small{}4.09} & {\small{}2.61} & {\small{}4.99}\tabularnewline
 & {\small{}17\%} & {\small{}1.50} & {\small{}1.46} & {\small{}1.70} & {\small{}2.26} & {\small{}1.89} & {\small{}3.22} & {\small{}2.10} & {\small{}3.75}\tabularnewline
 & {\small{}2.5\%} & {\small{}1.22} & {\small{}1.29} & {\small{}1.39} & {\small{}1.97} & {\small{}1.52} & {\small{}2.76} & {\small{}1.71} & {\small{}3.37}\tabularnewline
\hline 
\multirow{5}{*}{{\small{}Fig. \ref{Fig4}B}} & {\small{}97.5\%} & {\small{}2.49} & {\small{}2.82} & {\small{}2.61} & {\small{}3.82} & {\small{}2.68} & {\small{}4.71} & {\small{}2.98} & {\small{}5.76}\tabularnewline
 & {\small{}83\%} & {\small{}2.13} & {\small{}2.32} & {\small{}2.45} & {\small{}3.53} & {\small{}2.52} & {\small{}4.48} & {\small{}2.83} & {\small{}5.43}\tabularnewline
 & {\small{}50\%} & {\small{}1.81} & {\small{}1.84} & {\small{}2.04} & {\small{}2.69} & {\small{}2.14} & {\small{}3.64} & {\small{}2.47} & {\small{}4.63}\tabularnewline
 & {\small{}17\%} & {\small{}1.43} & {\small{}1.43} & {\small{}1.62} & {\small{}2.24} & {\small{}1.77} & {\small{}3.20} & {\small{}1.98} & {\small{}3.76}\tabularnewline
 & {\small{}2.5\%} & {\small{}1.20} & {\small{}1.26} & {\small{}1.37} & {\small{}1.98} & {\small{}1.49} & {\small{}2.87} & {\small{}1.77} & {\small{}3.63}\tabularnewline
\hline 
\multirow{5}{*}{{\small{}Fig. \ref{Fig4}C}} & {\small{}97.5\%} & {\small{}2.81} & {\small{}2.91} & {\small{}3.20} & {\small{}4.27} & {\small{}3.58} & {\small{}6.04} & {\small{}3.79} & {\small{}7.07}\tabularnewline
 & {\small{}83\%} & {\small{}2.76} & {\small{}2.89} & {\small{}3.16} & {\small{}4.22} & {\small{}3.54} & {\small{}6.00} & {\small{}3.72} & {\small{}6.99}\tabularnewline
 & {\small{}50\%} & {\small{}2.25} & {\small{}2.47} & {\small{}2.58} & {\small{}3.62} & {\small{}2.74} & {\small{}4.85} & {\small{}3.01} & {\small{}5.97}\tabularnewline
 & {\small{}17\%} & {\small{}1.95} & {\small{}2.20} & {\small{}2.13} & {\small{}3.21} & {\small{}2.25} & {\small{}4.20} & {\small{}2.64} & {\small{}5.45}\tabularnewline
 & {\small{}2.5\%} & {\small{}1.72} & {\small{}1.87} & {\small{}1.98} & {\small{}2.88} & {\small{}2.05} & {\small{}4.14} & {\small{}2.32} & {\small{}4.98}\tabularnewline
\hline 
\multirow{5}{*}{{\small{}Fig. \ref{Fig4}D}} & {\small{}97.5\%} & {\small{}2.49} & {\small{}2.82} & {\small{}2.62} & {\small{}3.83} & {\small{}2.68} & {\small{}4.72} & {\small{}2.99} & {\small{}5.77}\tabularnewline
 & {\small{}83\%} & {\small{}2.36} & {\small{}2.68} & {\small{}2.47} & {\small{}3.64} & {\small{}2.53} & {\small{}4.56} & {\small{}2.91} & {\small{}5.51}\tabularnewline
 & {\small{}50\%} & {\small{}2.06} & {\small{}2.11} & {\small{}2.34} & {\small{}3.29} & {\small{}2.45} & {\small{}4.11} & {\small{}2.73} & {\small{}5.19}\tabularnewline
 & {\small{}17\%} & {\small{}1.80} & {\small{}1.80} & {\small{}1.99} & {\small{}2.62} & {\small{}2.15} & {\small{}3.74} & {\small{}2.45} & {\small{}4.54}\tabularnewline
 & {\small{}2.5\%} & {\small{}1.64} & {\small{}1.56} & {\small{}1.59} & {\small{}2.25} & {\small{}2.13} & {\small{}3.53} & {\small{}2.33} & {\small{}3.95}\tabularnewline
\hline 
\multirow{5}{*}{{\small{}Fig. \ref{Fig4}E}} & {\small{}97.5\%} & {\small{}1.95} & {\small{}1.98} & {\small{}2.25} & {\small{}2.83} & {\small{}2.44} & {\small{}3.89} & {\small{}2.58} & {\small{}4.63}\tabularnewline
 & {\small{}83\%} & {\small{}1.68} & {\small{}1.69} & {\small{}1.84} & {\small{}2.46} & {\small{}2.11} & {\small{}3.38} & {\small{}2.19} & {\small{}4.00}\tabularnewline
 & {\small{}50\%} & {\small{}1.50} & {\small{}1.46} & {\small{}1.77} & {\small{}2.28} & {\small{}1.92} & {\small{}3.23} & {\small{}2.10} & {\small{}3.74}\tabularnewline
 & {\small{}17\%} & {\small{}1.28} & {\small{}1.33} & {\small{}1.55} & {\small{}2.11} & {\small{}1.61} & {\small{}2.94} & {\small{}1.85} & {\small{}3.59}\tabularnewline
 & {\small{}2.5\%} & {\small{}1.20} & {\small{}1.26} & {\small{}1.36} & {\small{}1.96} & {\small{}1.48} & {\small{}2.70} & {\small{}1.64} & {\small{}3.12}\tabularnewline
\hline 
\end{tabular}\caption{Percentile statistics similar to the boxplot statistics depicted in
Figure \ref{Fig4}.}
\label{Tab3}
\end{table}

For the SSP2-4.5 case, the 2.5-4.0 ECS GCMs project a likely (or 66\%)
2080--2100 warming ranging between 2.24 and 3.53 °C with median 2.69
°C, which well corresponds to the 2.2-3.4 °C (with median 2.7 °C)
range estimated by the \citet{CAT} as the likely warming that will
occur by the end of the 21st century on the basis of real-world current
policies.

Figure \ref{Fig4} shows that the reported ranges vary greatly according
to the adopted SSP scenario, and the selection of models used for
the simulations. In general, the ensemble of all simulations (apart
from those referring to the SSP1-2.6 scenario) appears to easily exceed
the 2.0 °C (safe) threshold suggested by \citet{Gao} by as early
as 2050. This would be true whether one chooses to use all CMIP6 GCM
simulations or the selection of them that corresponds to the IPCC
likely ECS range (between 2.5 and 4.0 °C). Only the SSP1-2.6 scenario
would ensure an average warming of roughly 1.8 °C {[}1.26-2.82 °C{]}
by 2080--2100 still using the GCMs corresponding to the IPCC likely
ECS-range and could satisfy the Paris Agreement warming target. The
SSP1-2.6 scenario is rather aggressive since it requires us to take
the ``green road'' of achieving a net-zero global CO\textsubscript{2}
emission condition as soon as 2050, which makes such policy extremely
expensive and potentially harmful for society.

However, Figure \ref{Fig4} and Table \ref{Tab3} suggest that in
the eventuality that only the simulations from the low-ECS macro-CGM
are considered, also the moderate SSP2-4.5 scenario could be rather
feasible for preventing a climatic crisis because by 2050 the average
warming will still be below 2.0 °C (roughly 1.77 °C {[}1.36-2.25 °C{]})
and by 2080--2100 it will rise to just 2.28 °C {[}1.96-2.83 °C{]},
instead of 2.69 °C {[}1.98-3.82 °C{]} under the IPCC's assumptions.
SSP2-4.5 is a “middle of the road” scenario where CO\textsubscript{2}
emissions stay about the same as today until the middle of the century,
when they begin to progressively fall but, even by 2100, they do not
reach net-zero. Thus, this SSP scenario requires no discernible changes
in the historical trends of socioeconomic elements for the next 2-3
decades. Development and income growth are uneven, and sustainability
progress is gradual. Adaptation policies are assumed to be adequate
to mitigate major climate-related hazards.

The climate change projection ensembles used here primarily come from
KNMI Climate Explorer's GCM average simulations for each SSP. A critique
could be formulated that the entire ensemble of individual simulations,
rather than just the ensemble average simulation or a single member
simulation for each model and SSP, should be employed, resulting in
larger projection ensembles. The topic is controversial and, in any
case, impossible to assess because the number of individual runs for
each model varies significantly among models. \citet{Scafetta_CliDyn_2022}
for example, emphasized that, because physical models must be both
accurate and precise, the main prediction of the models is their average
simulation, whereas the error-range from such average should be considered
a user requirement rather than a GCM property due to its actual internal
variability. The user's acceptable error-range for each model can
be bounded to the temperature variability within the decadal-bidecadal
scales because the models are supposed to reproduce the longer scales,
and this variability was estimated to be of the order of about $\pm0.1$
°C \citep[Appendix]{Scafetta_CliDyn_2022}. Such error can be statistically
ignored because it is significantly smaller than the difference observed
among the various GCM average simulations. In fact, also \citet{Hausfather_2022}
evaluated the projection ensembles using one simulation for each model
and SSP, which is what I did here.

\section{Re-assessment: adoption of optimized empirical climate models}

If the actual global warming from 1980 to 2022 is compatible with
that provided by UAH-MSU (or NOAA-STAR v. 5.0) lower troposphere temperature
data \citep{Spencer_2017,Zou_2023}, the actual impacts and risks
connected with predicted twenty-first-century climate changes are
much smaller, and the CMIP6 GCM simulations should not be used to
make policy. In such instance, all CMIP6 GCMs would be inadequate
for accurately estimating climate changes in the 21st century because
the actual ECS would be roughly one-third lower than even the low-ECS
macro-GCM, that is between 1 and 2 °C \citep[cf.][]{McKitrick_2020,Scafetta_2013,Scafetta_2023c,Stefani},
which would rule out nearly all CMIP6 GCMs.

There are not enough models that span the ECS range from 1 to 2 °C.
However, the case could be approximated by empirically scaling the
CMIP6 macro-GCM simulations to best reflect the warming documented
by satellite data from 1980 to 2022. The purpose is to produce hypothetical
GCM simulations that best hindcast the available data, so that their
twenty-first-century projections could be trusted for policy.

The proposed scaling algorithm works in first approximation because
on a global scale the performance of a GCM is mainly determined by
its ECS and TCR values. Moreover, it is easy to calculate that scaling
down the medium and high-ECS macro-GCMs to mimic the warming of the
low-ECS macro-GCM from 1980--1990 to 2011--2022, the three macro-GCM
ensembles would roughly overlap throughout the 21st century. In fact,
by using the data from Table \ref{Tab2}, the ratio between the 2011--2022
medians of the medium- and high-ECS ensembles and the low-ECS ensemble
are $0.80/0.62\approx1.3$ and $0.97/0.62\approx1.6$, respectively;
similar ratios are found using the data from Table \ref{Tab3}. By
using the 2080--2100 medians: for SSP1-2.6 the ratios are $2.1/1.5\approx1.4$
and $2.5/1.5\approx1.7$; for SSP2-4.5 the ratios are $3.3/2.3\approx1.4$
and $3.6/2.3\approx1.6$; for SSP3-7.0 the ratios are $4.1/3.2\approx1.3$
and $4.8/3.2\approx1.5$; and for SSP5-8.5 the ratios are $5.2/3.7\approx1.4$
and $6.0/3.7\approx1.6$, respectively. These ratios remain approximately
the same for all other percentile levels.

As a result, because the performance of one macro-GCM can be approximated
by an appropriate linear scaling of another one, it is legitimate
to modify the climate projection ensembles of the available macro-GCMs
to generate plausible climate projection ensembles from hypothetical
macro-GCMs by scaling them to best reproduce the available data. Another
advantage of the proposed methodology is that it can be applied to
a wide range of circumstances where the existing GCMs are unsatisfactory.
For example, the three proposed macro-GCMs can be scaled to recreate
both global surface and lower troposphere temperatures, or they can
be scaled and integrated with climate models that empirically reconstruct
natural cycles and extra solar components that the CMIP GCMs do not
reproduce, as \citet{Scafetta_2013,Scafetta_2021} already proposed.

Let us now address these two situations in detail.

\subsection{GCM scaling on the surface and lower troposphere temperature records}

As explained in Section 2, climate-change impacts and risks for the
21st century are assessed by the IPCC \citep{IPCC_2022a} using the
CMIP6 GCMs simulations under different SSP scenarios. However, \citet{Hausfather_2020}
suggested that the SSP2-4.5 scenario is the most plausible. According
to \citet{Pielke_2022}, only the SSP2-3.4 and SSP2-4.5 scenarios
should be deemed realistic since their emission trajectories are the
most compatible with recent history and present reference estimates
of the expected evolution of the global energy system over the next
three decades, but here only the SSP2-4.5 GCM simulations are available.
Thus, it is here suggested that the Hist+SSP2-4.5 scenario simulations
from the CMIP6 GCMs could serve as a basis for realistic climate change
projections for the 21st century.

Moreover, according to the analysis presented in Sections 2 and 4,
the best-performing subset of CMIP6 GCMs appears to be the one with
ECS values between 1.5 and 3.0 °C \citep{Lewis_2023,Scafetta_CliDyn_2022}.
The low-ECS macro GCM is composed of 15 GCMs. From this list it might
be possible to exclude the FGOALS-f3-L model and to add the BCC-CSM2-MR
($ECS=3.04$ °C) and MRI-ESM2-0 ($ECS=3.15$ °C) models. FGOALS-f3-L
may be excluded because, despite its ECS is on the border limit ($3.0$
°C), its TCR appears to be too high (1.94 °C). Instead, the other
two models might be included since, despite having an ECS slightly
higher than 3.0 °C, their TCR is low (1.72 and 1.64 °C, respectively).
In fact, a low-TCR macro-GCM would have a TCR equal to or less than
1.8 °C. All the other CMIP6 GCMs should be avoided for climate-change
policy since they are characterized by too high ECS or TCR values
and, as an ensemble, they overestimate the warming observed from 1980
to 2022.

In any instance, the optimal selection of GCMs that must be chosen
might be considered arbitrary. I propose an alternative methodology
that considers all available simulations. The ensemble projections
of each of the three macro-GCMs can be empirically scaled to best
fit the observed warming from 1980--1990 to 2012--2022 to ensure
an ideal outcome compatible with the observed warming during the same
period (and add a little bit more dispersion associated with the internal
variability of the models). In this study, the HadCRUT5 and UAH-MSU
lt v. 6.0 temperature records are deemed representative of the surface-based
and satellite-based global surface temperature estimations.

The necessary scaling factors are obtained from the median values
reported in Table \ref{Tab2} for each of the three ECS macro-GCMs,
as well as the warming of 0.605 °C and 0.432 °C shown by the HadCRUT5
and UAH-MSU lt v. 6.0 records from 2012 to 2022 relative to 1980--1990,
respectively. Thus, we have:
\begin{itemize}
\item Case \#1 uses HadCRUT5 --- the Hist+SSP2-4.5 GCM simulations of the
high-ECS group are scaled by the factor $0.605/0.97=0.62$, those
of the medium-ECS group are scaled by the factor $0.605/0.80=0.77$,
and those of the low-ECS group are scaled by the factor $0.605/0.620=0.98$;
\item Case \#2 uses UAH-MSU lt v. 6.0 --- the Hist+SSP2-4.5 GCM simulations
of the high-ECS group are scaled by the factor $0.432/0.97=0.45$,
those of the medium-ECS group are scaled by the factor $0.432/0.80=0.54$,
and those of the low-ECS group are scaled by the factor $0.432/0.620=0.70$.
\end{itemize}
The above ratios are characterized by a statistical relative error
of about $\pm$ 5\%.

In Case \#1, the climate change projection ranges for the 21st century
do not differ much from those determined only from the low-ECS macro
GCM simulations mentioned in Section 4 and reported in Table \ref{Tab3}
because the scaling factor in such a case is close to one. Case \#2,
on the other hand, necessitates a 30\% reduction in the climate simulations
of the low-ECS macro GCM. The {[}2.5\%, 17\%, 50\%, 83\%, 97.5\%{]}
percentiles of the simulated warming distributions from 1850--1900
covering each decade from 2000 to 2100 are shown in Table \ref{Tab4}.

\begin{table}[!t]
\centering{}%
\begin{tabular}{|c|c|cccccccccc|}
\hline 
\multicolumn{1}{|c}{} &  & \multicolumn{10}{c|}{Temperature change from 1850--1900 (°C)}\tabularnewline
\multicolumn{1}{|c}{} & ﻿Percentile & 2005 & 2015 & 2025 & 2035 & 2045 & 2055 & 2065 & 2075 & 2085 & 2095\tabularnewline
\hline 
 & 97.5\% & ﻿1.11 & 1.30 & 1.59 & 1.85 & 2.06 & 2.36 & 2.57 & 2.75 & 2.96 & 3.09\tabularnewline
Case \#1 & 83.0\% & 0.95 & 1.14 & 1.41 & 1.66 & 1.89 & 2.13 & 2.35 & 2.55 & 2.70 & 2.82\tabularnewline
Fig. \ref{Fig5}A & 50.0\% & 0.74 & 0.90 & 1.15 & 1.35 & 1.59 & 1.80 & 1.97 & 2.10 & 2.22 & 2.34\tabularnewline
(HadCRUT5) & 17.0\% & 0.48 & 0.65 & 0.86 & 1.04 & 1.23 & 1.43 & 1.58 & 1.72 & 1.86 & 1.94\tabularnewline
 & 2.5\% & 0.30 & 0.52 & 0.71 & 0.90 & 1.05 & 1.24 & 1.38 & 1.51 & 1.62 & 1.68\tabularnewline
\hline 
 & 97.5\% & 0.80 & 0.93 & 1.13 & 1.32 & 1.47 & 1.68 & 1.84 & 1.96 & 2.11 & 2.21\tabularnewline
Case \#2 & 83.0\% & 0.67 & 0.82 & 1.01 & 1.19 & 1.34 & 1.52 & 1.68 & 1.82 & 1.92 & 2.02\tabularnewline
Fig. \ref{Fig5}B & 50.0\% & 0.53 & 0.64 & 0.82 & 0.96 & 1.13 & 1.29 & 1.41 & 1.50 & 1.59 & 1.67\tabularnewline
(UAH-MSU lt v. 6.0) & 17.0\% & 0.34 & 0.46 & 0.62 & 0.74 & 0.87 & 1.03 & 1.13 & 1.23 & 1.33 & 1.39\tabularnewline
 & 2.5\% & 0.22 & 0.37 & 0.51 & 0.65 & 0.75 & 0.89 & 0.99 & 1.07 & 1.16 & 1.20\tabularnewline
\hline 
 & 97.5\% & 1.18 & 1.26 & 1.34 & 1.50 & 1.67 & 1.98 & 2.25 & 2.38 & 2.52 & 2.52\tabularnewline
Case \#3 & 83.0\% & 1.00 & 1.10 & 1.20 & 1.30 & 1.51 & 1.75 & 2.02 & 2.17 & 2.26 & 2.26\tabularnewline
Fig. \ref{Fig7}A & 50.0\% & 0.80 & 0.85 & 0.94 & 1.00 & 1.21 & 1.42 & 1.64 & 1.73 & 1.78 & 1.77\tabularnewline
(HadCRUT4.6) & 17.0\% & 0.54 & 0.60 & 0.65 & 0.69 & 0.85 & 1.06 & 1.25 & 1.34 & 1.41 & 1.37\tabularnewline
 & 2.5\% & 0.37 & 0.47 & 0.51 & 0.55 & 0.68 & 0.87 & 1.05 & 1.13 & 1.18 & 1.11\tabularnewline
\hline 
\end{tabular}\caption{Percentiles {[}2.5\%, 17\%, 50\%, 83\%, 97.5\%{]} of the distributions
of the best estimated projected warming from 1850--1900 to the 11-year
periods covering each decade from 2000 to 2100. The curves are depicted
in Figures \ref{Fig5} and \ref{Fig7}. Modeling}
\label{Tab4}
\end{table}

Due to the used scaling factors, the warming hindcasted by the models
from 1850--1900 to 1980--1990 may not match the observed warming.
This is unimportant because past data are characterized by increasing
uncertainty \citep{Morice_2021}. Furthermore, if surface temperature
records from 1980 to 2022 had to be scaled down to match the satellite-based
records, the real warming from the pre-industrial period to the present
would be less than what was reported \citep[cf.][]{Connolly2023,Scafetta_CliDyn_2021}.

\begin{figure}[!t]
\centering{}\includegraphics[width=1\textwidth]{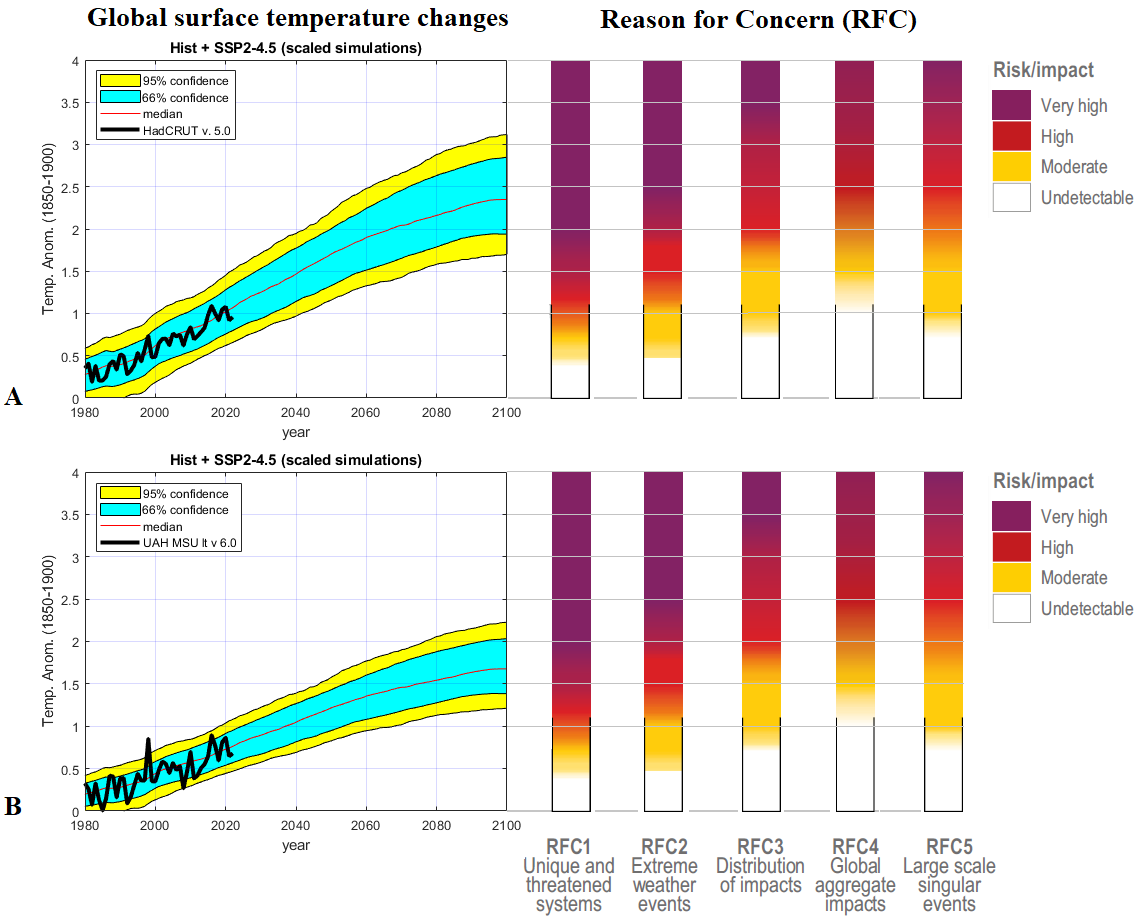}\caption{{[}A{]} Estimated global temperature changes using the Hist+SSP2-4.5
scenario and scaling the CMIP6 GCM simulations to optimally hindcast
the HadCRUT5 global surface temperature warming from 1980--1990 to
2012--2022. {[}B{]} The same as in {[}A{]} using CMIP6 GCM simulations
scaled to optimally hindcast the UAH-MSU lt v. 6.0 temperature warming
from 1980--1990 to 2012--2022. Both temperature records are baselined
with the modeled hindcast mean from 1980 to 1990. Table \ref{Tab4}
tabulates the depicted percentile ranges. {[}Right{]} Burning ember
diagrams (in function of the global temperature warming) of the main
five global reason for concern (RFC) assuming low to no adaptation
reported by the IPCC AR6 \citep{IPCC_2022a}.}
\label{Fig5}
\end{figure}

Figure \ref{Fig5}A's left panel depicts Case \#1. It shows the estimated
global temperature changes from 1980 to 2100, as well as their percentile
--- likely (66\%) and very likely (95\%) --- ranges, based on the
Hist+SSP2-4.5 scenario and scaling the CMIP6 GCM simulations to best
hindcast the warming shown by the HadCRUT5 global surface temperature
record from 1980--1990 to 2012--2022. The temperature record is
baselined with the modeled hindcast mean from 1980 to 1990.

Figure \ref{Fig5}B's left panel depicts Case \#2. It is the same
as in Figure \ref{Fig5}A, but the CMIP6 GCM simulations are now scaled
to best hindcast the warming seen in the UAH-MSU lt v. 6.0 temperature
record from 1980-1990 to 2012-2022. This scenario should be used if
it is demonstrated that non-climatic warm biases influence global
surface temperature records and that the satellite lower troposphere
temperatures better represent the actual global surface warming from
1980 to 2022. The temperature record is baselined with the modeled
hindcast mean from 1980 to 1990.

The displayed percentile ranges of the estimated global temperature
changes shown in Figures \ref{Fig5}A and \ref{Fig5}B are reported
in Table \ref{Tab4} for each decade from 2000 to 2100. The right
panels of Figure \ref{Fig5} depict the burning ember diagrams (in
function of the global temperature warming) of the main five global
reason for concern (RFC) assuming low to no adaptation as reported
in the IPCC AR6 \citep{IPCC_2022a}. The five RFCs are described as:
\begin{itemize}
\item RFC1 (unique and threatened systems) --- ecological and human systems
with limited geographic ranges due to climate-related factors and
high endemism or other distinguishing characteristics such as coral
reefs, the Arctic and its indigenous peoples, mountain glaciers, and
biodiversity hotspots;
\item RFC2 (extreme weather events) --- impacts and risks to human health,
livelihoods, assets and ecosystems from extreme weather events such
as heatwaves, heavy rain, drought and associated wildfires, and coastal
flooding;
\item RFC3 (distribution of impacts) --- impacts and risks that disproportionately
affect specific groups as a result of uneven distribution of physical
climate change hazards, exposure, or vulnerability;
\item RFC4 (global aggregate impacts) --- impacts on socio-ecological systems
that may be aggregated globally into a single metric, such as monetary
damages, lives lost, species extinction, or global ecosystem deterioration;
\item RFC5 (large-scale singular events) --- relatively significant, rapid,
and potentially permanent changes in systems driven by global warming,
such as ice sheet collapse or thermohaline circulation slowdown.
\end{itemize}
Figure \ref{Fig5}A suggests that the impacts and risks of climate
changes for all five RFCs could be relatively moderate (yellow-orange
flag) until 2050, and they will gradually increase until 2100. Only
some unique and particularly critically threatened environmental systems
may be at higher risk. In fact, Table \ref{Tab4} shows that the very
likely (95\% confidence) warming should range 1.15-2.21 °C (median
1.70 °C) by 2040--2060, and 1.65-3.03 °C (median 2.28 °C) by 2080--2100.

Figure \ref{Fig5}B shows the 21st-century climate projection depicted
in Figure \ref{Fig5}A is reduced by roughly 30\% to fit the lower
troposphere temperature records: the 95\% confidence warming ranges
from 0.82 to 1.58 °C (median 1.21 °C) by 2040--2060, and from 1.18
to 2.16 °C (median 1.63 °C) by 2080--2100. These temperature projections
are much lower than the IPCC AR6 estimates. For example, simply using
the GCMs with ECS between 2.5 and 4.0 °C, which is the IPCC likely
ECS-range, the global warming is estimated to be between 1.37 and
2.61 °C by 2040--2060 and 1.98-3.82 °C by 2080--2100 (Table \ref{Tab3}).

Cases \#1 and \#2 imply that the impacts and risks of climate change
could be moderate (yellow-orange flag) in all five RFCs until 2100
also using the SSP2-4.5 scenario.

\subsection{GCM scaling assuming natural oscillations}

\citet{Scafetta2010,Scafetta_2013,Scafetta_Climate_2021} proposed
empirical models for global climate change based on the evidence that
climate records appear to be characterized by several oscillations
that could be related to solar or astronomical harmonics. From a physical
point of view it might be possible that cloud formation, which the
GCMs partially parameterize with some free parameters that are carefully
tuned \citep{Mauritsen,Mignot}, may also strongly depend on some
non-radiative forcing (cosmic rays, interplanetary dust, etc.) modulated
by solar magnetic activity \citep{ScafettaMilani,Scafetta_2023c,Svensmark_2016,Svensmark_2022},
which the state-of-the-art climate models ignore. In fact, the GCMs
can also be directly tuned just to obtain an improved match to the
instrumental temperature record by changing their free parameters
\citep{Mauritsen(2020)}, which leaves the possibility of missing
mechanisms \citep[cf.][]{Kaufman}.

The proposed model assumes that the warming indicated by global surface
temperature records (e.g., HadCRUT3 and HadCRUT4) is sufficiently
accurate. It is made of a multi-harmonic natural component simulating
the hypothesized and empirically-derived astronomical/solar effect
superimposed on an empirically estimated anthropogenic-plus-volcanic
effect scaled from the GCM simulations. The empirical model also includes
a quasi 9.1-year oscillation that appears to be caused by solar-lunar
tidal forcings. \citet{Scafetta2010,Scafetta2012,Scafetta_2013} estimated
that if such natural solar/astronomical harmonic contribution is considered,
the climate's sensitivity to anthropogenic and volcanic forcings should
be approximately halved. This means that the empirically determined
anthropogenic-plus-volcanic effect can be approximated by multiplying
the ensemble average of all GCM simulations by a factor of about 0.5.
For more information, see \citet{Scafetta_2013,Scafetta_2023c}. The
underlying hypothesis is that the harmonics that have characterized
climate change in recent decades, centuries, and millennia will continue
to do so in the 21st century.

\begin{figure}[!t]
\centering{}\includegraphics[width=1\textwidth]{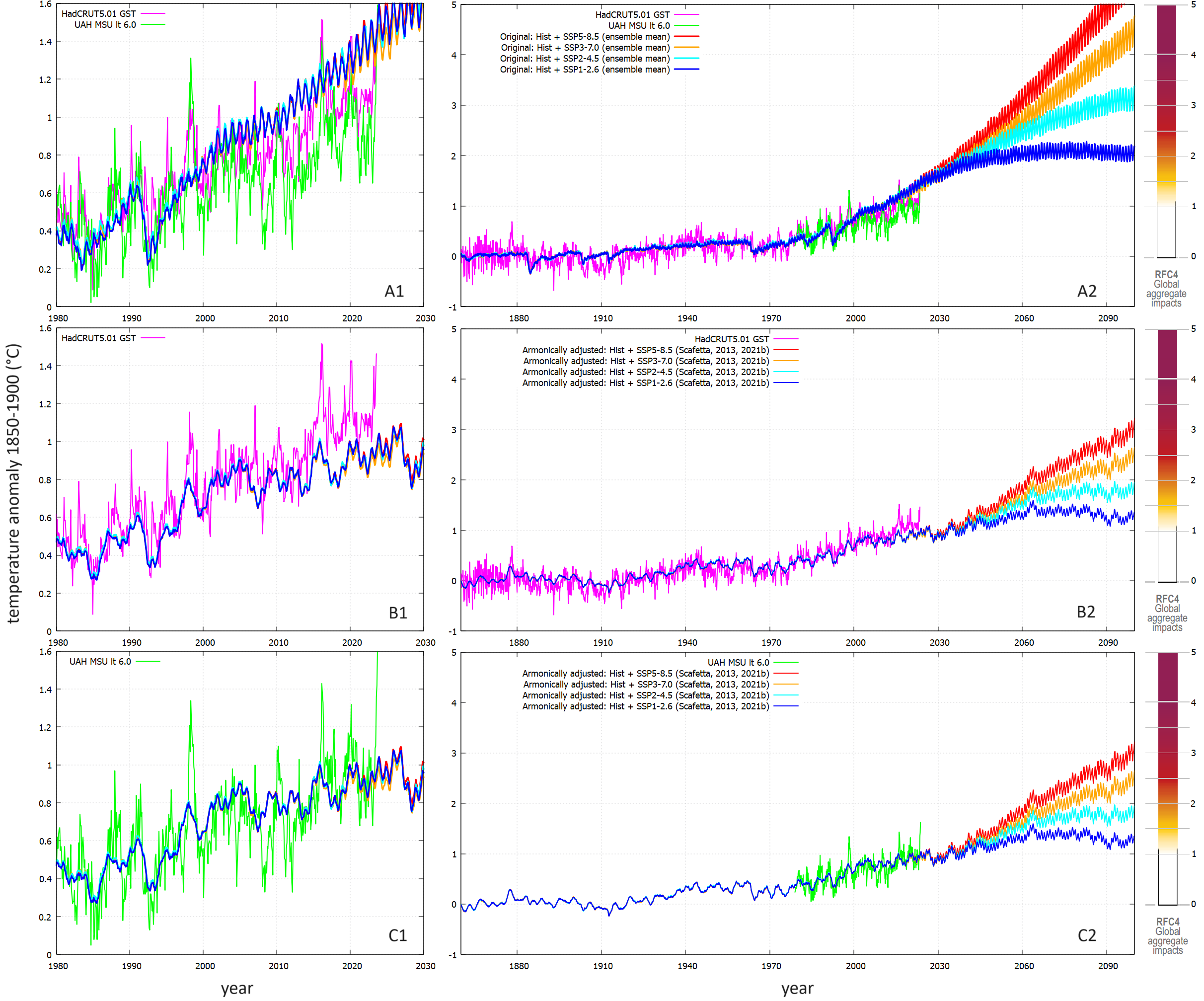}\caption{Comparison of the climate simulations and projections with their relative
global aggregated impacts and risks obtained with (A) the ensemble
average CMIP6 GCM simulations and (B and C) the harmonic empirical
global climate modeling proposed by \citet{Scafetta2010,Scafetta_2013,Scafetta_2021}.
The temperature data used are the (infilled) HadCRUT v.5 \citep{Morice_2021}
and the UAH-MSU lt v. 6.0 lower troposphere global temperature records
\citep{Spencer_2017}.}
\label{Fig6}
\end{figure}

Figure \ref{Fig6} compares the original CMIP6 GCM ensemble average
simulations to the proposed empirical model. The latter is based on
\citet{Scafetta_2021}'s 13-harmonics set (representing the solar-astronomical
induced variability of the climate system), which adds 7 interannual
identified harmonics to \citet{Scafetta_2013}'s original six harmonics
model covering the decadal-to-millennial scales. The model is completed
by an empirically derived anthropogenic-plus-volcanic signal, which
is obtained by halving the ensemble average of the GCM simulations,
implying that the actual ECS could range between 0.9 and 2.8 °C, as
recently evaluated by independent studies \citep{Lewis_2023,Scafetta_2023,Spencer2023}.

The proposed equation for the empirical global surface temperature
model is

\begin{equation}
T(t)=T_{0}+\sum_{i=1}^{13}A_{i}\sin\left[2\pi\left(f_{i}\cdot(t-2000)+\alpha_{i}\right)\right]+0.5\cdot GCM(t),\label{eq:1}
\end{equation}
where the coefficients $A_{i}$, $f_{i}$ and $\alpha_{i}$ per $i=1,...,13$
are reported in Table \ref{Tab5}, $GCM(t)$ is the ensemble average
of the GCM simulations for each SSP scenarios (which were downloaded
from KNMI Climate Explorer), and $T_{0}$ is the reference baseline
of the temperature anomalies that can be chosen, for example, to make
$T_{1850-1900}=0$. The harmonic coefficients were derived from astronomical
considerations and optimized for best data fitting using a Montecarlo
approach \citep{Scafetta_2021}. Throughout the Holocene, the period
of the quasi millennial cycle was theoretically and empirically estimated
to be about 983 years \citep{Scafetta2012b}, but $f_{1}$ in Eq.
\ref{eq:1} corresponds to a period of 760 years that was chosen to
simulate the skewness of the millennial cycle, which could be modeled
with the addition of the quasi 2318-year Hallstatt solar cycle \citep{ScafettaOrtolani,Scafetta2020}.
Indeed, paleoclimate temperature reconstructions from the last millennium
typically show a minimum around 1680 (corresponding to the coldest
period of the Little Ice Age during the Maunder solar minimum) and
a millennial maximum around 1077; the next solar millennial grad-maximum
is expected around 2060. \citet[figure 8]{Scafetta2012b} showed that
the proposed semi-empirical model well hindcast the quasi-millennial
cycle observed in climate records throughout the Holocene despite
both its phase and period were deduced from solar-astronomical considerations
alone. \citet[figure 5B]{Scafetta2012} validated an early version
of the same harmonic model by independently calibrating its decadal
and multi-decadal harmonics for the years 1850--1950 and 1950--2011,
demonstrating that the two independent climate hindcasts closely coincided
from 1850 to 2050 and optimally reproduced the global surface temperature
record from 1850 to 2011. \citet[figures 11 and 14]{Scafetta_2021}
also used the proposed model to reproduce temperatures since the Medieval
period, and \citet[figures 15]{Scafetta_2021} demonstrated that the
model, which was calibrated using data up to 2014, could also forecast
the warm periods that occurred in 2015--2016 and around 2020, as
well as the approaching 2023--2024 warm period.

\begin{table}[!t]
\centering{}%
\begin{tabular}{|c|ccc|}
\hline 
$i$ & $A_{i}$ (°C) & $f_{i}$ (1/y) & $\alpha_{i}$\tabularnewline
\hline 
1 & 0.3228 & 0.001318 & 0.1711\tabularnewline
2 & 0.05843 & 0.008696 & 0.4239\tabularnewline
3 & 0.08583 & 0.01639 & 0.1516\tabularnewline
4 & 0.03339 & 0.0500 & 0.1482\tabularnewline
5 & 0.02407 & 0.09615 & 0.0198\tabularnewline
6 & 0.02651 & 0.1075 & 0.4973\tabularnewline
7 & 0.02163 & 0.1340 & 0.7107\tabularnewline
8 & 0.02721 & 0.1666 & 0.6167\tabularnewline
9 & 0.02598 & 0.1909 & 0.4086\tabularnewline
10 & 0.03257 & 0.2086 & 0.9312\tabularnewline
11 & 0.02758 & 0.2752 & 0.7666\tabularnewline
12 & 0.02537 & 0.2812 & 0.7917\tabularnewline
13 & 0.02472 & 0.3480 & 0.9746\tabularnewline
\hline 
\end{tabular}\caption{Coefficients for the harmonic component of the empirical climate model
given by Eq. \ref{eq:1}.}
\label{Tab5}
\end{table}

The left panels of Figure \ref{Fig6} zoom in on the right panels
between 1980 and 2030 to better show the models' performance over
the last 50 years. The temperature data used here are the (infilled)
HadCRUT v.5 \citep{Morice_2021}, which is chosen as an average representative
of the most recent global surface temperature records \citep{Scafetta_CliDyn_2022},
and the UAH-MSU lt v 6.0 global temperature record by \citet{Spencer_2017},
which is nearly identical to the NOAA STAR v. 5.0 lt record \citep{Zou_2023}.
The lower troposphere temperature records are adopted as representative
of the real surface temperature change since 1980 if the surface records
are influenced by non-climatic biases \citep{Scafetta_CliDyn_2021,Connolly_2021,Connolly2023,Soon2023}.
Finally, the picture depicts global aggregated impact estimates based
on the burning ember diagram proposed by the IPCC AR6 \citep{IPCC_2022a}.
The data represent temperature anomalies relative to the pre-industrial
period of 1850--1900.

Figure \ref{Fig6}A depicts the original GCM simulations. Only the
net-zero emission SSP1-2.6 scenario ensures that global surface temperatures
do not significantly exceed the 2 °C (safe) threshold \citep{Gao,Tol_2015},
which satisfies to the Paris Agreement warming targets. However, as
shown in \ref{Fig6}-A1, these average models are running too hot
in relation to both global surface and lower troposphere temperature
records.

Figure \ref{Fig6}B compares the four SSP average simulations generated
by the proposed empirical model of Eq. \ref{eq:1} to the HadCRUT5
record. The figure shows that, in addition to the (net-zero emission)
SSP1-2.6 scenario, also the moderate SSP2-4.5 scenario would ensure
a global surface temperature of less than 2 °C throughout the 21st
century, with an average of about 1.8 °C by 2080--2100. Furthermore,
the extreme (and highly unlikely) SSP5-8.5 scenario is projected to
warm the global climate on average by only 3 °C relative to pre-industrial
levels, which is only moderately concerning. However, as illustrated
in \ref{Fig6}-B1, the empirical model appears to underestimate the
HadCRUT5-measured global surface temperature warming. Yet, it agrees
well with the HadCRUT3 and HadCRUT4 records \citep{Scafetta_2013,Scafetta_2021};
therefore, the current discrepancy with the (infilled) HadCRUT5 appears
to be attributable to the adjustments made to the latter, which conceal
the 2000--2014 temperature ``hiatus'' or ``pause'' that is still
revealed by the satellite temperature records.

In fact, Figure \ref{Fig6}C is like Figure \ref{Fig6}B, except the
simulations are now compared to the lower troposphere temperature
record from UAH-MSU. The empirical model \citep[first proposed by][]{Scafetta2010,Scafetta_2013}
appears to have accurately predicted the observed warming. As a result,
if satellite temperature data better represent the actual global surface
warming from 1980 to 2022, the model's empirical calibration can be
assumed validated, potentially making its future climate projections
trustworthy as well.

\begin{figure}[!t]
\centering{}\includegraphics[width=1\textwidth]{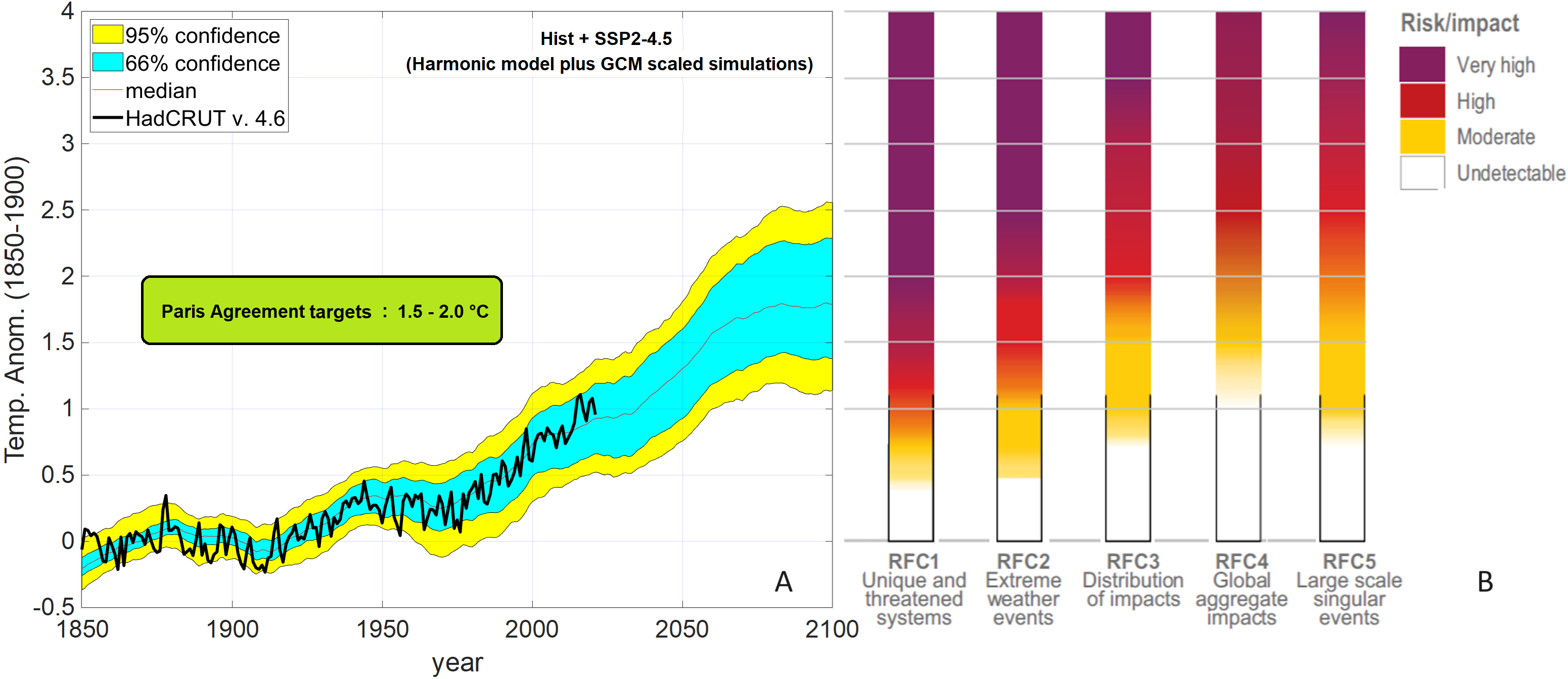}\caption{(A) The harmonic empirical global climate model with the SSP2-4.5
scenario, against the HadCRUT4.6 record (1850--2021) \citep{Morice2012}.
(B) Burning ember diagrams (in function of the global temperature
warming) of the main five global reason for concern (RFC) assuming
low to no adaptation reported by the IPCC AR6 \citep{IPCC_2022a}.
Table \ref{Tab4} tabulates the depicted percentile ranges from 2000
to 2100.}
\label{Fig7}
\end{figure}

Figure \ref{Fig7}A compares the harmonic empirical global climate
model with the SSP2-4.5 scenario to the HadCRUT4.6 record \citep{Morice2012},
which is used here as a compromise between the (infilled) HadCRUT5
and the lower troposphere satellite temperature records. \citet{Scafetta_2021}
calibrated his empirical model until 2014 using the HadCRUT4.6 record.
The model projection range is equivalent to Figure \ref{Fig5}A. Figure
\ref{Fig7}A also shows that the proposed harmonic empirical model
reconstructs the multidecadal modulation (consisting of quasi 60-year
oscillations forming a sequence of quasi-30-year warming or cooling
periods) observed in the temperature record since 1850 much better
than the GCM simulations, which show only monotonically increasing
patterns interrupted by sporadic volcano eruptions (see Figures \ref{Fig1}
and \ref{Fig6}-A2); the same quasi-60-year modulation can be reconstructed
with appropriate total solar activity records, as demonstrated by
\citet{Scafetta_2023c}.

By extending the same natural modulation up to 2080--2100, the empirical
climate projection predicts that the temperature will range from 1.15
to 2.52 °C (median 1.78 °C), which roughly matches the 1.26-2.82 °C
range produced by the SSP1-2.6 scenario using GCMs with ECS ranging
from 2.5 to 4.5 °C as reported by the IPCC. Table \ref{Tab4} reports
the depicted percentile ranges from 2000 to 2100.

The warming rate is expected to remain relatively low from 2000 to
2035 and from 2065 to the end of the century because the empirical
model predicts negative phases of the Atlantic Multidecadal Oscillation
(AMO) during these periods, whereas the warming rate is expected to
increase from 2035 to 2065 when the AMO is forecasted to be positive.
The AMO highlights a well-known large quasi-60-year oscillation of
the climate system that is not predicted by GCMs \citep{Scafetta_2013}
but has been observed for millennia \citep{Knudsen,Scafetta2014a,Wyatt_2013};
it is most likely of solar/astronomical origin, along with other larger
oscillations of the climate system \citep{Neff,Scafettaetal2013,Scafetta2014b,Scafetta2020,Scafetta_2023c,ScafettaMilani}.

\section{Summary of the results}

Figures \ref{Fig5} and \ref{Fig7} show three ``realistic'' global
warming projections for the 21st century. They imply that the impacts
and risks of realistic climatic changes are significantly lower than
what the IPCC expected, even when only the SPP2-4.5 scenario was offered.

\begin{figure}[!t]
\centering{}\includegraphics[width=1\textwidth]{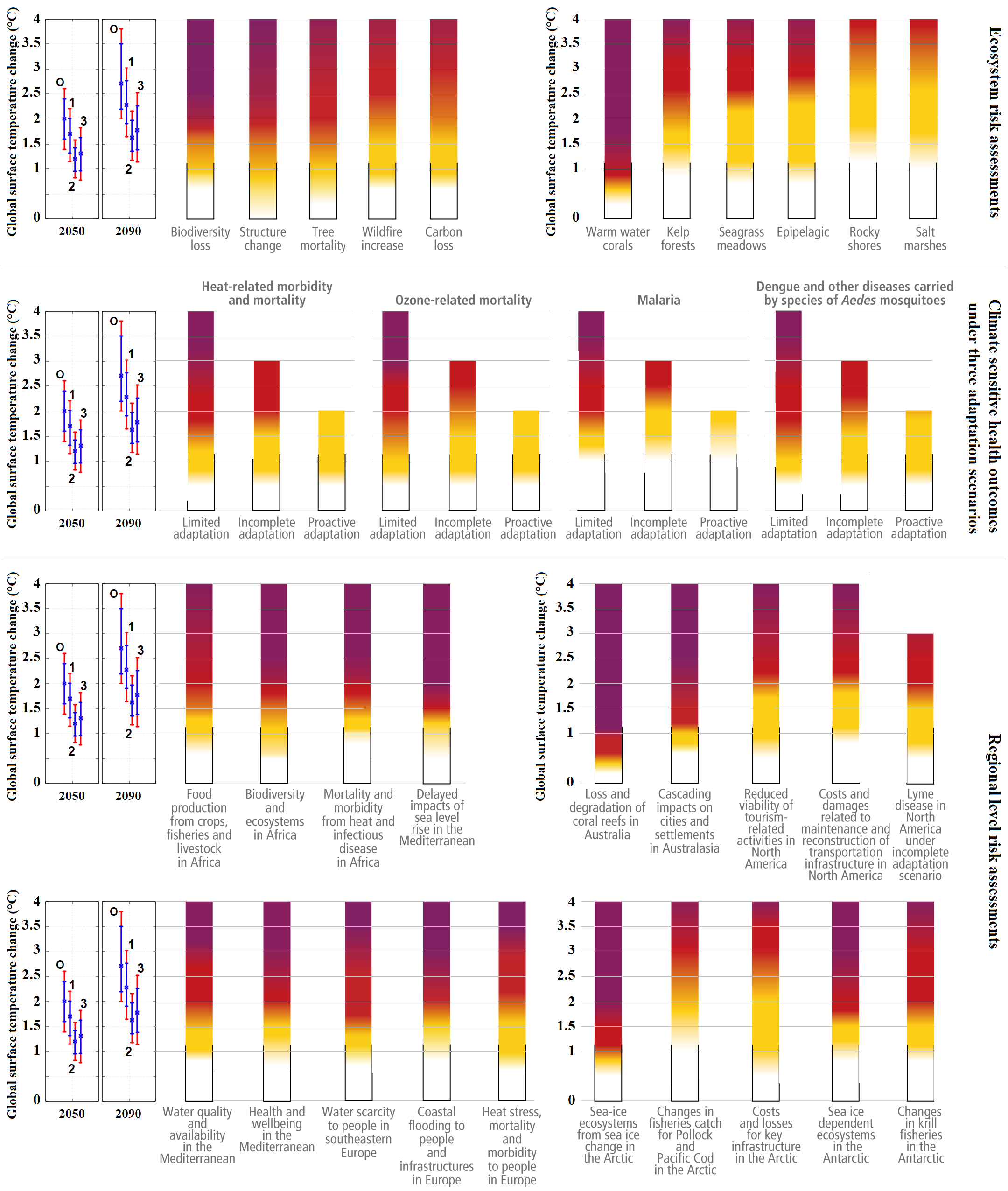}\caption{{[}Left{]} Estimated warming in 2040--2060 and 2080--2100 according
to the SSP2-4.5 scenario and four cases: (O-original) using the GCMs
with ECS between 2.5 and 4 °C (IPCC option); (1) using the GCMs scaled
on the low-ECS macro GCM and the HadCRUT5 record (Figure \ref{Fig5}A);
(2) using the GCMs scaled on the low-ECS macro GCM and the UAH-MSU
lt v. 6.0 record (Figure \ref{Fig5}B); (3) using the empirical harmonic
model (Eq. \ref{eq:1}) and the HadCRUT4.6 record (Figure \ref{Fig7}A).
{[}Right{]} Burning ember diagrams (in function of the global temperature
warming) representing the estimated global and local impacts and risks
of global warming for several ecosystems and health outcomes as reported
in the IPCC AR6 \citep[Summary for Policymakers]{IPCC_2022a}.}
\label{Fig8}
\end{figure}

The left panels of Figure \ref{Fig8} summarize the estimated warming
in 2040--2060 and 2080--2100 according to the SSP2-4.5 scenario
and the warming projected ranges at 66\% and 95\% confidence as deduced
in four cases (see Tables \ref{Tab3} and \ref{Tab4}):
\begin{itemize}
\item Ranges \#O are the original ones and derive from the GCMs with ECS
values between 2.5 and 4 °C, which are in line with the IPCC's prediction
of its likely ECS range;
\item Ranges \#1 derive from the GCMs scaled on the low-ECS macro GCM and
assumes the 1980--2022 warming of the HadCRUT5 record, as shown in
Figure \ref{Fig5}A;
\item Ranges \#2 derive from the GCMs scaled on the low-ECS macro GCM and
assumes the 1980--2022 warming of the UAH-MSU LT v. 6.0 record, as
shown in Figure \ref{Fig5}B.
\item Ranges \#3 derive from the empirical harmonic model (Eq. \ref{eq:1})
depicted in Figure \ref{Fig7}A.
\end{itemize}
The right panels of Figure \ref{Fig8} show a selection of the same
burning ember diagrams (in function of the global temperature warming)
of regional and global risk assessments relative to several ecosystems,
and disease-health situations adopted by the IPCC AR6. A detailed
explanation of each RFC is found in \citet{IPCC_2022a}. The first
line represents examples of ecosystem risk assessments; the second
line represents examples of climate sensitive health outcomes under
limited, incomplete, and proactive adaptation scenarios; the third
and fourth lines represents examples of regional level risk assessments.
Unless otherwise specified, the burning ember diagrams represent situations
in which little or no adaptation is planned. As more sophisticated
adaptation policies are implemented, the color scale shifts upward
in a way like what is shown in the second row of Figure \ref{Fig8}.

Most of the time, for cases \#1-3, the impacts and risks of climate
changes would be moderate (yellow-orange flag) until 2050. The SSP2-4.5
scenario may also ensure an average global surface temperature of
less than 2 °C by 2100, and hence also this moderate SSP should be
considered compatible with the Paris Agreement warming objective.
As a result, while suitable adaptation techniques may always be required,
they may be rather affordable because the impacts and risks of actual
future climate change are expected to be moderate.

\section{Discussion and conclusion}

The IPCC used the CMIP6 GCMs to assess the probable risks and impacts
of climate change on global and regional scales over the next century
\citep{IPCC_2021,IPCC_2022a}. Climate change estimates are influenced
by the climate's sensitivity to radiative forcing as well as by the
projected greenhouse gas emissions, which are linked to varying rates
and kinds of socioeconomic developments. However, there is a great
deal of uncertainty in both conditions that must be restricted to
properly assessing realistic climatic hazards for the 21st century
and developing appropriate climate policies to optimally address them.

The ECS of the CMIP6 GCMs ranges between 1.8 and 5.7 °C, but the IPCC
AR6 acknowledged the existence of a ``\emph{hot}'' model problem
and claimed that the actual ECS may likely range between 2.5 and 4.0
°C, with a best estimate of around 3.0 °C \citep{Sherwood_2020,Hausfather_2022}.
However, recent research suggests that the expected ECS range should
vary within lower values between 1.5 and 3 °C \citep{Nijsse_2020,Scafetta_GRL_2022,Scafetta_CliDyn_2022,Scafetta_2023,Lewis_2023,Spencer2023}.
Furthermore, according to a number of empirical studies, the actual
ECS values could even be significantly lower, ranging between 1 and
2 °C \citep{Bates,Lindzen,McKitrick_2020,Scafetta_2013,Scafetta_2023c,Stefani}.

The IPCC AR6 investigates a range of SSP scenarios for the 21st century
without assigning a probability to their plausibility. In any case,
despite the questionable visibility given to SSP5-8.5 (the worst-case
scenario), which yields the largest and most alarming projected global
warming of up to 4-8 °C (66\%) by 2080--2100, Table 12.12 of the
IPCC AR6 \citep[pp. 1856]{IPCC_2021} already reports for the entire
21st century low confidence in the direction of any change in the
frequency, severity or extent of frost, river floods, landslides,
aridity, hydrological drought, agricultural and ecological drought,
fire weather, mean wind speed, severe wind storms, tropical cyclones,
sand and dust storms, heavy snowfall and ice storms, hail, snow avalanche,
coastal floods, coastal erosion, marine heatwaves, air pollution weather
or radiation at earth’s surface. Medium and high confidence of changes
are mostly expected in climatic impact-driver types more directly
associated with increasing atmospheric CO\textsubscript{2} concentration
at surface and global warming such as increasing mean air temperature,
extreme heat, sea level, mean ocean temperature, ocean salinity and
ocean acidity; with decreasing cold spell, snow, glacier and ice sheet,
permafrost, lake, river and sea ice, and dissolved oxygen; mean precipitation
will increase in some regions and decrease in others.

However, recent research argued that the alarming SSP3-7.0 and SSP5-8.5
scenarios are \emph{likely} and \emph{very likely} unrealistic, respectively
\citep{Burgess_2020,Hausfather_2020,Pielke_2021a,Pielke_2021b}. These
studies indicate that the radiative forcing functions derived from
the SSP2-4.5 (or even SSP2-3.4) scenario are the most plausible. The
SSP2-4.5 is a moderate scenario; it projects about half of the 21st-century
warming than what the SSP5-8.5 scenario does (Figure \ref{Fig1})
and is thus far less alarming.

With the aforementioned factors in mind, it was proposed here to use
only the SSP2-4.5 scenario and the GCMs with ECS $\leq3$ °C to more
precisely assess ``realistic'' global and regional impacts and risks
that could be associated with climate changes that are expected to
occur in the 21st century, and to compare them with the Paris Agreement
warming target of keeping global surface temperature below 2 °C above
the pre-industrial levels throughout the 21st century. To optimize
the result even more, the simulation ensembles containing the low,
medium, and high-ECS macro-GCMs were linearly scaled to best reflect
the real global surface warming recorded from 1980--1990 to 2012--2022.

According to the IPCC, if there is little-to-no adaptation, the impacts
and risks of projected climate change will be moderate-high (orange-red
flag) by 2040--2060, and the situation might worsen considerably
by 2100 even if the SSP2-4.5 moderate scenario is implemented. In
fact, according to the analysis reported in Table \ref{Tab3}, the
GCMs within the IPCC's preferred ECS range of 2.5-4.0 °C project a
warming of 1.98-3.82 °C by 2080--2100. Thus, the IPCC \citep{IPCC_2018}
concludes that only net-zero-emission scenarios like the SSP1-2.6
(which could produce a warming of 1.26-2.82 °C by 2080--2100) should
be adopted to avoid too alarming climatic changes, which are expected
to begin if global surface temperatures rise more than 2-2.5 °C above
the 1850--1900 level in a few decades \citep{Tol_2015}. Climate-change
alarmism and world-wide proposals for prompt implementations of net-zero
emission policies are based on such claims.

However, using only the low-ECS models (ECS $\lessapprox3.0$ °C)
and the SSP2-4.5 scenario, Table \ref{Tab3} suggests that global
warming in the 21st century will be moderate, ranging from 1.36 to
2.25 °C (median 1.77 °C) by 2050 and from 1.96 to 2.83 °C (median
2.28 °C) by 2080--2100, which partially overlaps with the upper warmer
half of the climate projection obtained using the SSP1-2.6 scenario
and the GCMs with ECS of 2.5-4.0 °C. Thus, climate change impacts
and risks will worsen by the end of the 21st century, albeit at a
slower rate than predicted by the IPCC using the same SSP2-4.5 scenario.
As a result, the SSP2-4.5 scenario, which is moderate and affordable,
may be close enough to roughly meet the Paris Agreement warming target,
whereas the SSP2-3.4 scenario, which could be even more realistic
\citep{Pielke_2022}, should more likely fully meet it.

I also proposed an alternative methodology for estimating ``realistic''
21st-century climate projections and assessing their respective impacts
and risks. In fact, the low-ECS macro-GCM appears to be slightly warmer
than global surface temperature records and there are serious concerns
about the reliability of the global surface temperature records, which
cannot be ignored. In fact, their warming appears to be excessive
in comparison to alternative temperature records, such as satellite-based
ones relative to the lower troposphere \citep{Spencer_2017,Zou_2023},
and there are various evidences suggesting their contamination from
urban heat islands and other non-climatic surface factors \citep{Scafetta_CliDyn_2021,Scafetta_2023,Connolly_2021,Soon2023}.
There are also concerns regarding the ability of the GCMs to properly
reconstruct decadal to millennial natural climate oscillations \citep{Scafetta_2013,Scafetta_2021}.
As a result, all GCMs may be grossly inadequate for estimating climate
change in the 21st century, as also \citet{McKitrick_2020} concluded.
Thus, the models need to be corrected and upgraded with new relevant
physical mechanisms. It is possible to agree with \citet{McCarthy}
who recently showed the inability of the CMIP5 and CMIP6 GCMs in properly
hindcasting the Atlantic Meridional Overturning Circulation and concluded
``\emph{if these models cannot reproduce past variations, why should
we be so confident about their ability to predict the future?}''.

To address the above issues, I have proposed an alternative methodology
that uses empirical modifications of the actual GCM projection ensembles
via appropriate linear scaling in such a way to simulate the outputs
of hypothetical climate models that could accurately represent the
warming observed from 1980 to 2022. The 1980--2022 period was selected
because it is covered by a variety of temperature records with low
statistical errors. This methodology would essentially simulate hypothetical
GCMs that are supposed to optimally reproduce the data. Simple testing
validates the proposed methodology because scaling the projection
ensembles of the three macro-GCMs to a similar level from 1080--1990
to 2011--2022 results in projection ensembles that approximately
overlap throughout the 21st century.

The proposed methodology may also be justified by considering that
the GCMs are extremely sensitive to small modifications of their internal
free parameters, in particular to those regarding cloud formation,
and even GCM modelers adopt complex tuning approaches to explicitly
calibrating them to better match historical data \citep{Mauritsen(2020),Mignot}.
Section 4 proposes and investigates several of these modeling approaches,
the results of which are depicted in Figures \ref{Fig5} and \ref{Fig7}.

If the warming of the HadCRUT5 record from 1980--1990 to 2011--2022
is assumed correct, it is found that the SSP2-4.5 scenario produces
climate projections similar to those produced by the low-ECS macro-GCM.
In fact, the projected warming ranged from 1.65 to 3.03 °C by 2080--2100,
with a median of 2.28 °C (Table \ref{Tab4}, case \#1). This conclusion
is unsurprising given that the low-ECS macro-GCM has already successfully
recreated the HadCRUT5 warming.

However, if the reference warming is that reported by lower troposphere
satellite temperature data \citep{Spencer_2017,Zou_2023}, the warming
of the low-ECS macro-GCM simulations must be lowered by about 30\%.
As a result, global warming by 2080--2100 is projected to range from
1.18 to 2.16 °C (median 1.63 °C) above pre-industrial levels using
the SSP2-4.5 scenario (Table \ref{Tab4}, case \#2), which is well
below the (safe) threshold of 2.0 °C and is even cooler than the 1.26-2.82
°C estimate obtained with the GCMs with ECS within the IPCC likely
range of 2.5 and 4.0 °C using the SSP1-2.6 net-zero emission scenario.

A similar result was obtained with an empirical climate model that
assumes that the global surface temperature record is sufficiently
accurate, but also takes into account temperature changes caused by
empirically identified large climate cycles and/or solar effects that
the CMIP6 GCMs do not replicate \citep{Scafetta2010,Scafetta_2013,Scafetta_2021};
this case projects a warming ranging from 1.15 to 2.52 °C with median
1.78 °C by 2080--2100. Unfortunately, the IPCC ignores such semi-empirical
modeling of the climate system although it has been developed and
discussed in the scientific literature, and it should not be dismissed
lightly given that the GCMs fail to reproduce the natural oscillations
observed throughout the Holocene. They do not, for example, reproduce
any of the Holocene warm periods, such as the Roman and Medieval warm
periods, which indicate a quasi-millennial oscillation, a quasi-60-year
oscillation, and many other natural climate oscillations. Also this
kind of empirical modeling predicts very modest ECS values, ranging
at least between 1 and 3 °C, but more likely between 1 and 2 °C \citep{Bates,Lindzen,Scafetta_2013,Scafetta_2021,Scafetta_2023c,Stefani}.

In conclusion, as \citet{Hausfather_2020} pointed out, it is past
time to stop treating worst-case climate change scenarios (e.g., SSP3-7.0
and SSP5-8.5) as the most likely outcomes, because only realistic
and pragmatic scenarios, such as SSP2-4.5 or SSP2-3.4, can lead to
sound policies that can be accepted by all nations. Furthermore, net-zero
scenarios such as SSP1-2.6 look to be equally unattainable, as the
depletion of crucial metals required for low-carbon solar and wind
technologies, as well as electric vehicles and their chargers, appears
to make low-carbon technology production impossible on the very large
scale required to substitute fossil fuels \citep{Groves}. In fact,
despite the IPCC AR6 reports are rather alarming because global surface
temperatures were projected to rise by up to 4-8 °C above pre-industrial
levels according to alarming but unrealistic shared socioeconomic
pathways (see Figure \ref{Fig1} and \citealp{IPCC_2021}), with catastrophic
consequences in many situations \citep{IPCC_2022a}, Figures \ref{Fig5},
\ref{Fig7} and \ref{Fig8} show that ``realistic'' climate change
impacts and risks for the 21st century will likely be much more moderate
than what the IPCC claims. This is because there is a growing body
of evidence that the actual ECS may be rather low (1.5-3.0 °C, or
even 1-2 °C) for a variety of reasons derived from direct CMIP6 GCM
assessments, likely warming biases affecting global surface temperature
records, and a (likely solar induced) natural variability that the
current climate models do not reproduce. According to the semi-empirical
climate modeling proposed above, the climate system will likely warm
by less than 2.0-2.5 °C by 2080--2100, and on average less than 2.0
°C, also if the moderate SSP2-4.5 scenario is implemented. As a result,
rapid decarbonization and net-zero emission scenarios such as the
SSP1-2.6 are shown to be unnecessary to maintain global surface temperature
below 2 °C throughout the 21st century.

\begin{figure}[!t]
\centering{}\includegraphics[width=1\textwidth]{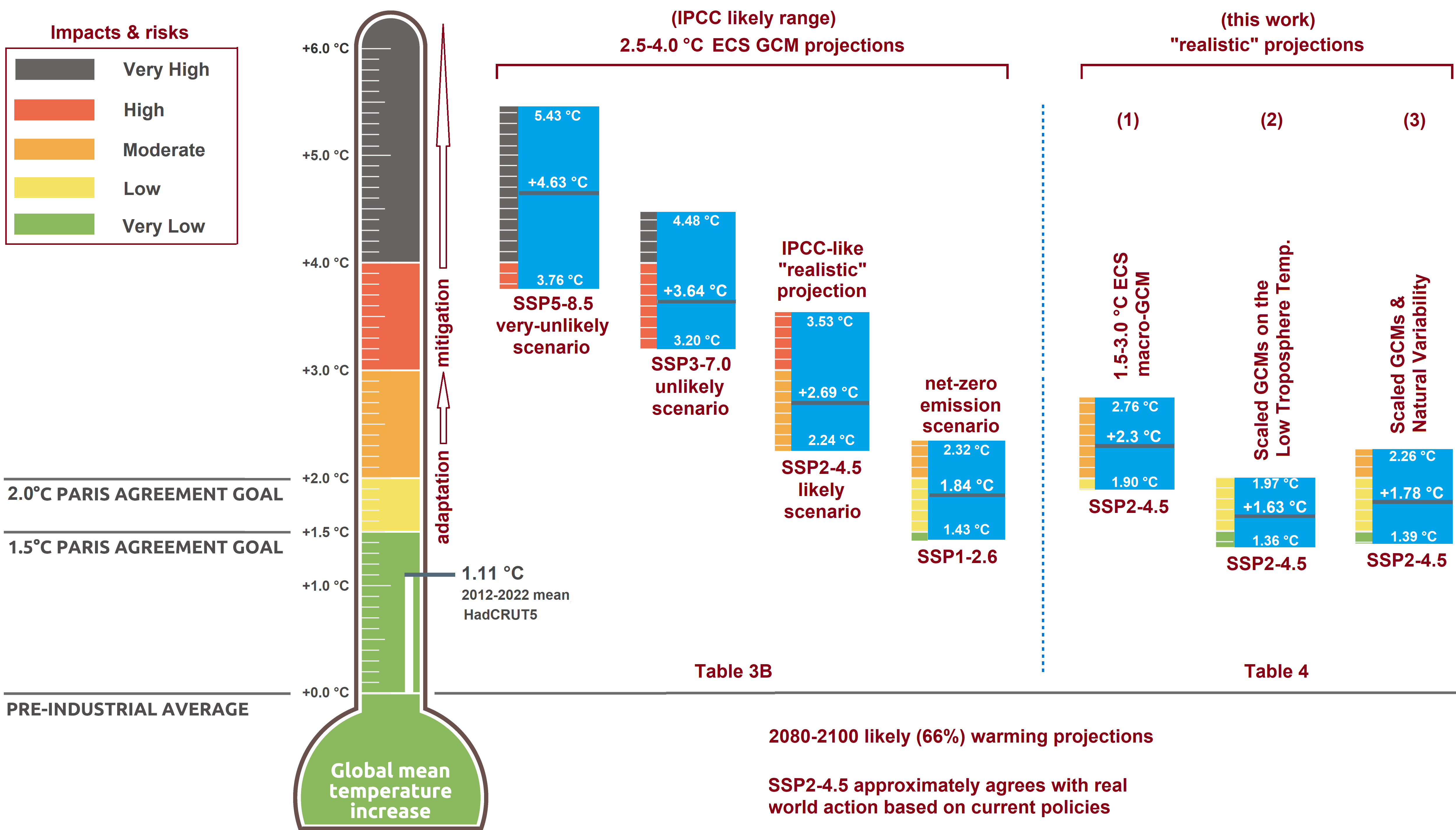}\caption{Summary and comparison of the impacts and risks of global warming
projections for the 2080--2100 period herein obtained (Tables \ref{Tab3}B
and \ref{Tab4}) versus the climate \textquotedblleft thermometer\textquotedblright{}
proposed by \citet{CAT}.}
\label{Fig9}
\end{figure}

Figure \ref{Fig9} employs the climate ``thermometer'' proposed
by \citet{CAT} to summarize the above findings by contrasting the
projections derived from the IPCC climate assumptions, where only
the SSP1-2.6 net-zero emission scenario could satisfy the 2.0 °C target,
with the new proposed assessments of ``realistic'' global warming
impacts and risks obtained using the three semi-empirical models discussed
above with the pragmatic SSP2-4.5 scenario that approximately agrees
with the real world action based on current policies (Tables \ref{Tab3}B
and \ref{Tab4}).

As a result, despite predictions that the climate system would continue
to warm throughout the 21st century, there is no compelling evidence
of an impending global disaster caused by manmade greenhouse gas emissions.
The 2.0 °C Paris-agreement warming target for the 21st century can
likely be met even under the feasible and moderate SSP2-4.5 emission
scenario because future climate change is expected to be modest enough
that any potential related hazards can be addressed efficiently through
effective and low-cost adaptation strategies, without the need for
implementing rapid, expensive, and technologically likely impossible
net-zero decarbonization policies.

\section*{Declarations}

\subsection*{Funding}

The author declares that no funds, grants, or other support were received
during the preparation of this manuscript.

\subsection*{Competing Interests}

The author has no relevant financial or non-financial interests to
disclose.

\subsection*{Author Contributions}

NS has performed the analysis and written the study.

\subsection*{Data availability}

The data used in the paper can be downloaded from (accessed on 01/10/2023):
\begin{itemize}
\item KNMI Climate Explorer: \href{https://climexp.knmi.nl/selectfield_cmip6.cgi}{https://climexp.knmi.nl/selectfield\_cmip6.cgi}
\item Supplementary data from \citet{Hausfather_2022}
\item CRUTEM5 Global Temperature: \href{https://www.metoffice.gov.uk/hadobs/crutem5/}{https://www.metoffice.gov.uk/hadobs/crutem5/}
\item CRUTEM4 Global Temperature: \href{https://www.metoffice.gov.uk/hadobs/crutem5/}{https://www.metoffice.gov.uk/hadobs/crutem5/}
\item UAH MSU v. 6.0 Global Temperature: \href{http://vortex.nsstc.uah.edu/data/msu/v6.0/tlt/uahncdc_lt_6.0.txt}{http://vortex.nsstc.uah.edu/data/msu/v6.0/tlt/uahncdc\_lt\_6.0.txt}
\end{itemize}

\section*{Appendix A. Supplementary data}

The supplementary data file contains the temperature simulations depicted
in Figures \ref{Fig1}, \ref{Fig5}, \ref{Fig6} and \ref{Fig7}.\\
\href{https://ars.els-cdn.com/content/image/1-s2.0-S1674987123002414-mmc1.xlsx}{https://ars.els-cdn.com/content/image/1-s2.0-S1674987123002414-mmc1.xlsx}

\newpage{}

\end{document}